\documentclass[journal]{IEEEtran}
\usepackage{amsmath,graphicx}
\usepackage{amssymb}
\usepackage{lineno,hyperref}
\usepackage{url}
\usepackage{float}
\usepackage{textcomp}
\usepackage{xcolor}
\usepackage{algorithm}
\usepackage{algpseudocode}
\usepackage{subcaption}
\usepackage{lipsum}
\usepackage{multirow}
\usepackage{enumitem}
\usepackage{cite}

\begin{document}

\title{Time-varying Graph Signal Estimation \\ via Dynamic Multi-hop Topologies}

\author{Yi~Yan,~\IEEEmembership{Student~Member,~IEEE,}~Fengfan~Zhao,~\IEEEmembership{Student~Member,~IEEE,}~and~Ercan~E.~Kuruoglu,~\IEEEmembership{Senior~Member,~IEEE}
	\thanks{Yi Yan, Fengfan Zhao, and Ercan E. Kuruoglu are affiliated with Tsinghua-Berkeley Shenzhen Institute, Tsinghua University, and Institute of Data and Information, Tsinghua Shenzhen International Graduate School. Corresponding author: Ercan E. Kuruoglu. This work is supported by Shenzhen Science and Technology Innovation Commission under Grant JCYJ20220530143002005, Shenzhen Ubiquitous Data Enabling Key Lab under Grant ZDSYS20220527171406015, and Tsinghua Shenzhen International Graduate School Start-up fund under Grant QD2022024C.}
}


\maketitle

\begin{abstract}
	The assumption of using a static graph to represent multivariate time-varying signals oversimplifies the complexity of modeling their interactions over time. We propose a Dynamic Multi-hop model that captures dynamic interactions among time-varying node signals, while also accounting for time-varying edge signals, by extracting latent edges through topological diffusion and edge pruning. The resulting graphs are time-varying and sparse, capturing key dynamic node interactions and representing signal diffusion to both near and distant neighbors over time.	The Dynamic Multi-hop Estimation algorithm is further proposed, accurately representing the interaction dynamics among node signals while enabling adaptive estimation of time-varying multivariate signals spatially and temporally. The Dynamic Multi-hop Estimation is evaluated under two real-world datasets of brain network and stock market for the online estimation of partially observed time-varying signals corrupted by noise.

	
\end{abstract}

\begin{IEEEkeywords}
	graph signal processing (GSP), dynamic graphs, time-varying signals, and graph learning.
\end{IEEEkeywords}

\IEEEpeerreviewmaketitle

\section{Introduction}
\IEEEPARstart{G}{raphs} Signal Processing (GSP) methodologies have demonstrated significant success in machine learning tasks such as classification, regression, and clustering on graph nodes, largely due to their ability to leverage the irregularities within graphs \cite{Ortega_graph_2018, Dong_Graph_ML_2020, Leus_2024_GSP_review}. 
Conventional GSP algorithms typically rely on graph shifts, which are defined based on the graph topology, and this topology is often assumed to be static. 
However, the signals on graph nodes may be static or time-varying.
Introducing time-varying data often requires additional techniques beyond standard GSP methods.

One approach to time-varying graph signals was the adaptive graph Least Mean Squares (GLMS) algorithm \cite{bib_LMS}, which is based on a $l_2$-norm optimization with the assumption of Gaussian noise and exploiting band-limited spectral graph filters to conduct online estimations of graph signals on sampled observations. 
The Graph Diffusion LMS (GDLMS) extends GLMS by replacing spectral graph filters with spatial diffusion \cite{Roula_2017_LMS_Diffusion}.  
In GSP, the Adaptive Graph-Sign-Diffusion (GSD) algorithm was recently proposed in which a sign diffusion was accomplished by defining a fixed binary update based on $l_1$-norm optimization methods and graph (spatial) convolution on graphs \cite{yan_2023_sign}. 
However, a common assumption among the previously mentioned adaptive GSP algorithms is that they operate on a known static graph: graph shifts, diffusion, and filters are predefined and time-invariant. 
This represents an oversimplification, particularly for handling time-varying signals.
Additionally, graph topology in datasets often undergoes sparsification and standardization, which may remove potentially crucial links between key nodes, leaving them missing \cite{lu2024latent}. 
In other words, several hidden or latent edges may not be explicitly represented in the provided graph topology \cite{de2022latent}. 
The lack of representation should be addressed by extracting the time-varying nature from the graph signals themselves.
It is necessary to approach the problem from an alternative perspective that allows for dynamic signals on the edges and potentially reveals the missing or hidden node interactions. 

Most previously discussed algorithms on time-varying graph signals assume the underlying graph topology is fixed whereas for time-varying tasks, a dynamic topology could potentially provide a more effective representation \cite{kazi2022differentiable, de2022latent}.
If the data interaction between nodes is dynamic, simply assigning static weights to these edges is insufficient to capture the dynamics and may lead to inferior performance at solving the targeted tasks \cite{lu2024latent, de2022latent}. 
An example of effective dynamic representation is found in pandemic prediction tasks in which the underlying graph topology is dynamic based on daily population mobility between regions \cite{Panagopoulos2021pandemic}.
Another example of representing time-varying interactions is a dynamic graph formed by thresholding a time-varying regression of gene activation that reflects one-to-one gene activation patterns over time \cite{Kuruoglu_2016_gene}. 
One last example of dynamic graphs can be found using topological parameters of dynamic financial complex networks to capture network structure characteristics in order to predict market crushes \cite{2024_Market_Qin}.
These methods rely on domain-specific knowledge, making them difficult to generalize, which highlights the need for a more generalized, data-driven model.


Pairwise relationship between two data elements within multi-variate data can be represented using various different approaches. 
For example, in Statistical Graphical Models (SGM), the correlation and the precision matrices are adopted to construct graph topologies; these statistical measurements are the relationships between pairs of variables, which is ideal for forming edge connections between two interacting variables as nodes \cite{lauritzen1996graphical}. 
Taking the Gaussian Graphical Model (GGM) as a specific example of the SGM, GGM assumes that the multivariate signals follow the multivariate Gaussian distribution, and edges between these variables are the conditional dependencies or non-zero partial correlation coefficients, which could be formulated as a conditional dependency structure in the form of a graph \cite{Altenbuchinger_2020_GMM, yuan2007_GMM_model, dong_2016_learning_laplacian}.
The forming of the graph shifts combined with the graph smoothness assumption on the multivariate signals gives us a topology based on Gaussian prior that is suitable for GSP tasks \cite{dong_2016_learning_laplacian}.
Given an underlying graph structure at a single time instance, assuming that the dynamics of a time-varying graph only have some minor changes between time instances, a small perturbation analysis of the Laplacian matrix can be done to model the time-evolution of the dynamic graph \cite{sardellitti_2021_online_small_pertubation}. 
Inference of a dynamic topology from time-varying data can be tackled from another angle on an iterative prediction-correction approach on a GGM \cite{natali2021online}. 
The causal relationships between data can be modeled using a VAR model, making it possible to infer a graph topology from the underlying time-series vector \cite{zaman2020online, Kuruoglu_2016_gene}. 
A dynamic graph can be formed from the correlated properties as well by considering the space-time interactions directly from the statistical measures in the underlying multivariate signal \cite{liu2019graph, liu2019spatiotemporal, yamada2019time}. 
These methods often depend on strong assumptions about data distributions or predefined model structures, which may not accurately capture the full complexity of real-world data or dynamic systems.

Since graph edges represent interactions between nodes, changes in these interactions and their strength are often unrecognized and underrepresented in static graphs, as static graphs imply static interactions. 
In the context of time-varying signals on static graphs, this can result in the loss of edges that represent crucial node interactions when constructing the static graph. 
Recent research has introduced a more generalized signal representation called Topological Signal Processing (TSP), which addresses edge signal representation by generalizing spatial and spectral operations from graph nodes to simplicial complexes using Hodge Laplacians \cite{Barbarossa_2020, Barbarossa_2020_SPM}. 
From a GSP perspective, techniques such as transformation, prediction, sampling, and regularization can be enhanced by incorporating signals on both nodes and edges, thus fully capturing the evolving interactions and dynamics over time.
However, a key challenge remains in capturing the dynamic nature of time-varying node signals, as static graphs fail to account for changing node interactions. 

In this paper, we propose the Dynamic Multi-hop model to capture changes in interactions of time-varying node signals through dynamic graphs.
Our Dynamic Multi-hop model addresses the limitations of static graph representations by providing a more comprehensive view of evolving graph structures in the context of time-varying signals. 
Dynamic Multi-hop considers not only the immediate 1-hop neighbor interactions (graph edges) of each node but also captures how signals dynamically diffuse to neighbors several edges away, represented as time-varying edge signals.
Time-varying signals on both graph nodes and edges are represented using TSP techniques. 
Pruning techniques based on SGM are applied to ensure our Dynamic Multi-hop model forms time-varying, sparse topologies in the spatial domain
In other words, Dynamic Multi-hop enhances the topology by introducing latent edges through topological diffusion and pruning based on metrics defined on time-varying weights, revealing hidden dynamics previously unrepresented by a fixed topology.
Our aim for Dynamic Multi-hop is to provide a more accurate representation of interaction dynamics among node signals, enabling the formation of a data model for further analysis of time-varying graph signals and enhancing the performance of existing GSP methodologies.
Leveraging the captured dynamics, improved performance on the task of online estimation of time-varying node signals is demonstrated in two real-world scenarios: brain network modeling and stock market analysis.
The contributions of this paper can be summarized as follows:
\begin{itemize}
	\item The Dynamic Multi-hop is a data-driven approach that augments static graphs that contain time-varying node signals by constructing additional underrepresented latent edges. 
	The Dynamic Multi-hop captures the hidden interactions among nodes, representing signal diffusion and propagation over time through edges that carry time-varying signals. 
    In addition, by incorporating pruning techniques, the resulting dynamic graphs formulated by Dynamic Multi-hop are ensured to maintain sparsity.
	
	\item Utilizing the captured dynamic of node signals in terms of the dynamic graphs and time-varying edge weights, a Dynamic Multi-hop estimation algorithm is made available for effective online estimation of time-varying signals on the nodes. 
	The resulting algorithm performs well in the challenging setting when partial observations of node signals are dynamic over time and corrupted by noise.
	The Dynamic Multi-hop Estimation is capable of denoising the noisy signal effectively, interpolating the missing observations accurately, and forecasting the next-step predictions adaptively.
\end{itemize}

The paper is organized as follows. 
Section~\ref{sec_back} gives the preliminaries for GSP. 
The Dynamic Multi-hop is derived in Section~\ref{section_Dynamic_Multi_Hop}.
In Section~\ref{sec_adaptive}, we further extend the Dynamic Multi-hop into adaptive estimations of time-varying signals. 
The experiment results and discussions on real data of brain network and stock market are given in Section~\ref{sec_brain} and in Section~\ref{sec_stock} respectively.

\begin{figure*}
	\centering
	\begin{subfigure}{0.45\textwidth}
		\centering
		\includegraphics[trim= 0 30 0 30,width=\textwidth]{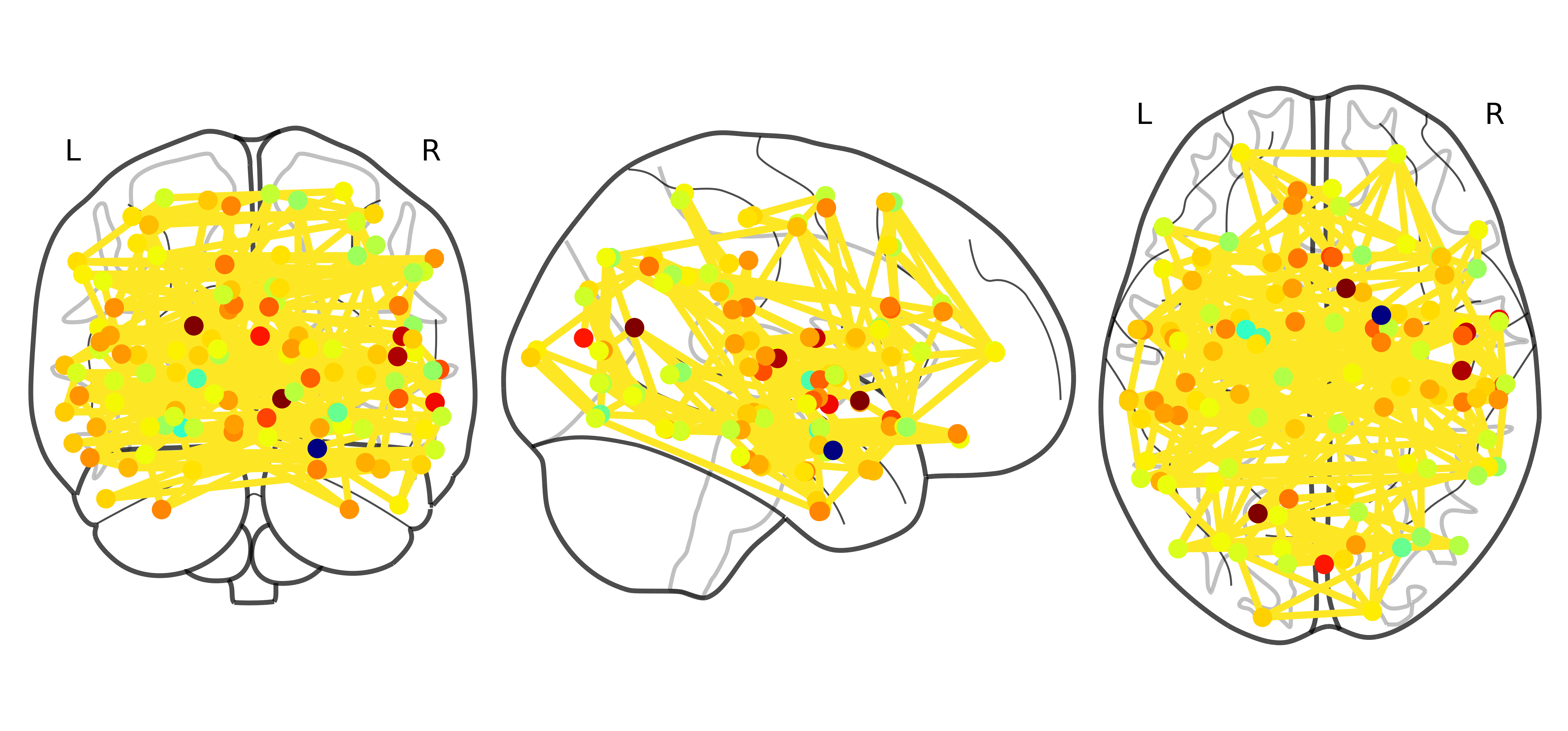}
		\caption*{Normal subject}
	\end{subfigure}
	\begin{subfigure}{0.45\textwidth}
		\centering
		\includegraphics[trim= 0 30 0 30,width=\textwidth]{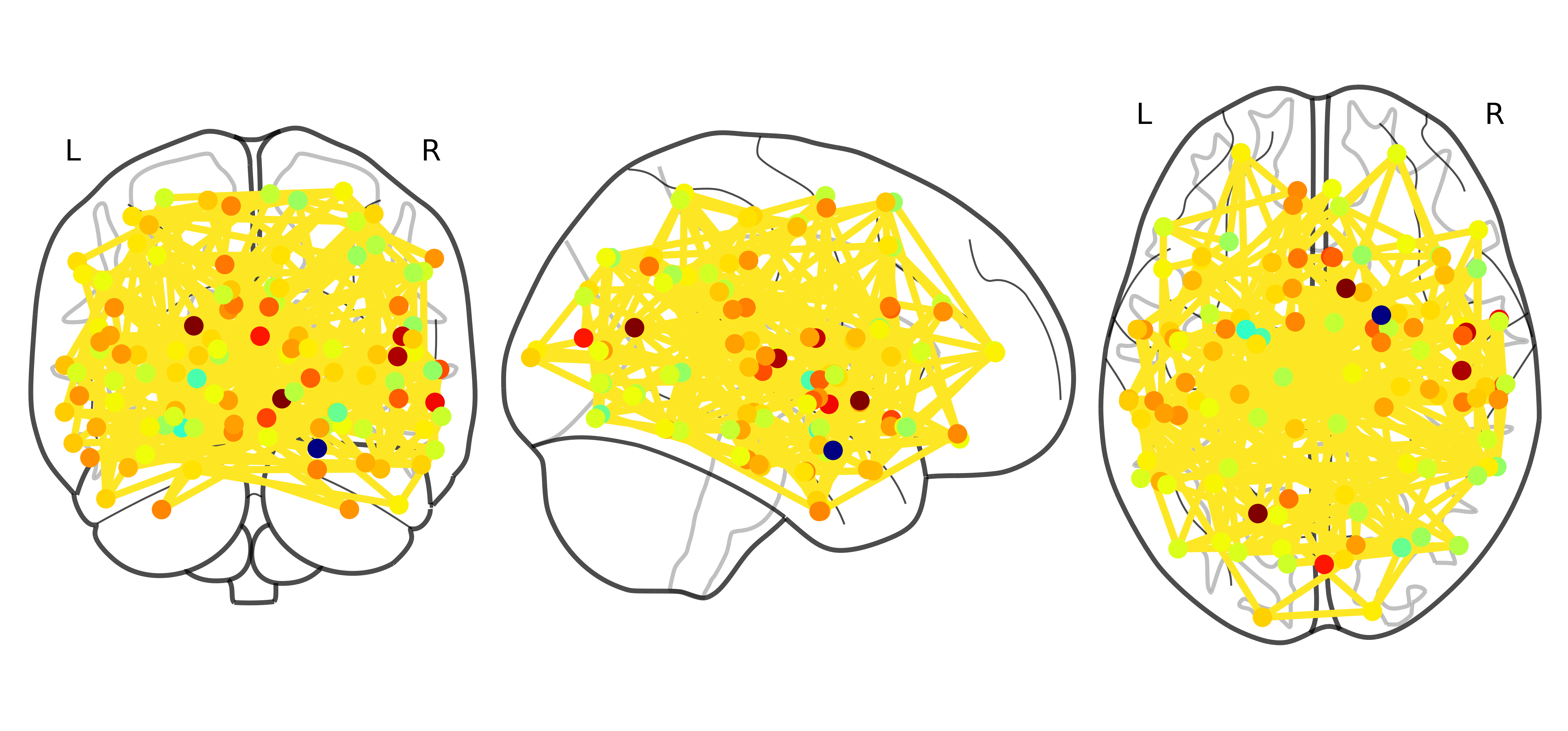}
		\caption*{Autistic subject}
	\end{subfigure}
	\caption{Static brain networks for normal and autistic subjects depicted using three different layouts. }
	\label{fig_brain_total_graph}
\end{figure*}

\section{Background}
\label{sec_back}
A graph $\mathcal{G}$ can be defined using three sets $\mathcal{V}$, $\mathcal{E}$ and $\mathcal{W}$, in which the set $\mathcal{V}$ is the set of $N_n$ nodes, the set $\mathcal{E} = {e_1 \dots e_{N_e}}$ is the set of $N_e$ edges, and $\mathcal{W} = {w_1 \dots w_{N_e}}$ are the weights of the edges respectively. 
The edge connectivity of $\mathcal{G}$ is recorded in the adjacency matrix $\mathbf{A}$, with the magnitude of $ij^{th}$ element being the edge weight for the specific edge connecting node $v_{i}$ to node $v_{j}$.
The degree matrix $\mathbf{D}$ is a diagonal matrix with the $i^{th}$ diagonal element known at the node degree of node $v_{i}$ and is computed as the sum of all the edge weights of the edges that connect to node $v_{i}$. 
A graph signal $\boldsymbol{x}$ is the function values defined on the nodes of the graph $\mathcal{G}$; each element in $\boldsymbol{x}$ corresponds to a signal at one single node. 
Without specifying otherwise, we will also refer to the graph signal as node signals and use these two terms interchangeably throughout this paper.
To make the Graph Fourier Transform (GFT) analogy in GSP, the graph Laplacian matrix $\mathbf{L}$ needs to be defined $\mathbf{L=D-A}$.
Then, GFT is accomplished using the eigenvector decomposition (EVD) of $\mathbf{L}$: $\mathbf{L=U\Lambda U}^{T}. $
Here the EVD results in $N_n$ distinctive pairs of eigenvalues and eigenvectors.  
The matrix $\mathbf{\Lambda}$ is the diagonal eigenvalue matrix with $N_n$ eigenvalues sorted in increasing order $\mathbf{\Lambda} = \text{diag}(\mathbf{\lambda}) = \text{diag}(\text{vec}(\lambda_1 \dots \lambda_{N_n}))$, where vec() is the vectorize operation. 
The matrix $\mathbf{U}$ is composed of the $N_n$ orthonormal eigenvectors of $\mathbf{L}$ \cite{DiLorenzo2018_sampling}. 
The GFT $\boldsymbol{s}=\mathbf{U}^{T}\boldsymbol{x}$ transforms the projection of $\boldsymbol{x}$ onto $\mathbf{U}$, giving us the spectral domain signal $\boldsymbol{s}$.
The inverse GFT $\mathbf{\boldsymbol{x}=U\boldsymbol{s}}$ transforms $\boldsymbol{s}$ from $\boldsymbol{s}$ spectral-domain to spatial domain. 

A band-limited graph signal $\boldsymbol{x}_{0}$ has the property that it only has a limited set of frequency components $\mathcal{F}$ with $|\mathcal{F}| < N_n$ in the spectral domain. 
To define a band-limited filter $\bf{\Sigma}_{\mathcal{F}}$, we can represent $\mathbf{\Sigma}_{\mathcal{F}}$ as diagonal matrix and a function of $\mathbf{\Lambda}$:	$\mathbf{\Sigma}_{\mathcal{F}} = \text{diag}(h(\lambda_1) \dots h(\lambda_{N_n}))$,  where $h(\lambda_i) = 1$ if $\lambda_i \in \mathcal{F} \quad \forall \, \lambda_i \in \mathbf{\lambda}$ and $0$ otherwise.
Notice that low-pass, high-pass, or band-pass filters on graphs are special cases of band-limited filters and can be designed based on modifying $\mathcal{F}$ \cite{bib_GSP_filter_design}. 
In GSP, a common assumption is that the graph signals to be processed are smooth \cite{Ortega_graph_2018}.  
In order to promote the smoothness of graph signals, one can apply a low-pass filter $\mathbf{\Sigma}_l$ in the spectral domain to the graph signal. 

The EVD can become unstable as the size of the graph topology increases \cite{Shuman_2011_Chebyshev}.
Moreover, each change in topology requires a newly defined graph Laplacian $\mathbf{L}[t]$ to reflect the topological change, meaning that EVD will be repeated to formulate the GFT. 
Luckily, the spectral graph filters can be expressed alternatively using spatial techniques.
By approximating the spectral filters using polynomials that directly operate the graph signal $\boldsymbol{x}$ using the spatial topological connections, the limitations caused by the EVD are circumvented:
\begin{equation}
	\mathcal{C}(\mathbf{L}, \mathcal{F})\boldsymbol{x} = \mathbf{U}\mathbf{\Sigma}_\mathcal{F}\mathbf{U}^T\boldsymbol{x} \approx \sum_{p=0}^{P} \theta_p \mathbf{L}^p \boldsymbol{x}, 
	\label{eq_L_convolution}
\end{equation}
where $P$ is the number of polynomials and $\theta_{p}$ is the weight of the $p^{th}$ polynomial.
The solution on the right side of the approximation in \eqref{eq_L_convolution} is obtained from algebraic manipulations of a series of shifted Chebyshev polynomials shown in \cite{Shuman_2011_Chebyshev}. 
For example, suppose we use the \eqref{eq_L_convolution} to express a low-pass filtering, then the weights $\theta_p$ are designed to mimic low-pass behavior $\mathbf{U}\mathbf{\Sigma}_l\mathbf{U}^T\boldsymbol{x} \approx \sum_{p=0}^{P} \theta_p \mathbf{L}^p \boldsymbol{x}$.
Similar concepts can be extended to band-limited filters or high-pass filters. 

Suppose the graph signal is represented using only a few select sampled nodes in the spatial domain. In that case, this sparse representation can be systematically obtained based on a designated sampled or observation set $\mathcal{S}\subseteq\mathcal{V}$. 
When the set $\mathcal{S}$ is obtained by sampling, the design strategies are usually to maximize a selection of properties of the graph signal $\boldsymbol{x}$ by a reduction of the number of elements in the graph signals from $N$ to $|\mathcal{S}|<N$ \cite{DiLorenzo2018_sampling}. 
Once the sampling set is finalized, we can record it into a diagonal sampling or masking matrix $\mathbf{M}$: the $i^{th}$ diagonal entry of $\mathbf{M}$ are equal to 1 when the signal on node $v_{i}$ is sampled in $S\subseteq\mathcal{V}$ and $0$  when the signal on node $v_{i}$ is is not sampled. 
It is worth pointing out that sampling matrix  $\mathbf{M}$ has the properties of being idempotent and self-adjoint.

\section{Dynamic Multi-hop Topologies}		\label{section_Dynamic_Multi_Hop}
\begin{figure}
	\centering
	\begin{subfigure}{0.45\textwidth}
		\centering
		\includegraphics[width=\textwidth]{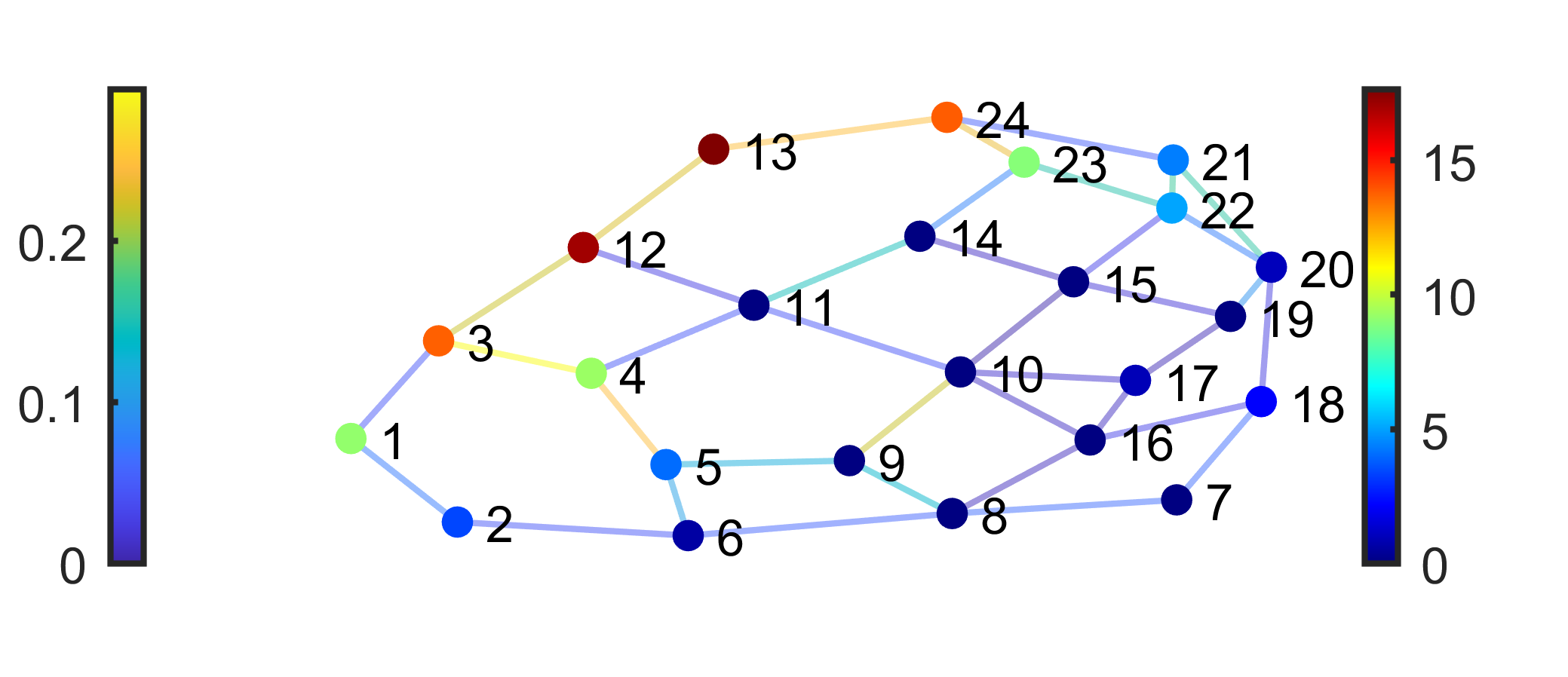}
		\vspace{-30pt}
		\caption{original $\mathcal{G}$}
		\label{fig_original_synthetic}
	\end{subfigure}
	\hspace{-10 pt}
	\begin{subfigure}{0.45\textwidth}
		\centering
		\includegraphics[width=\textwidth]{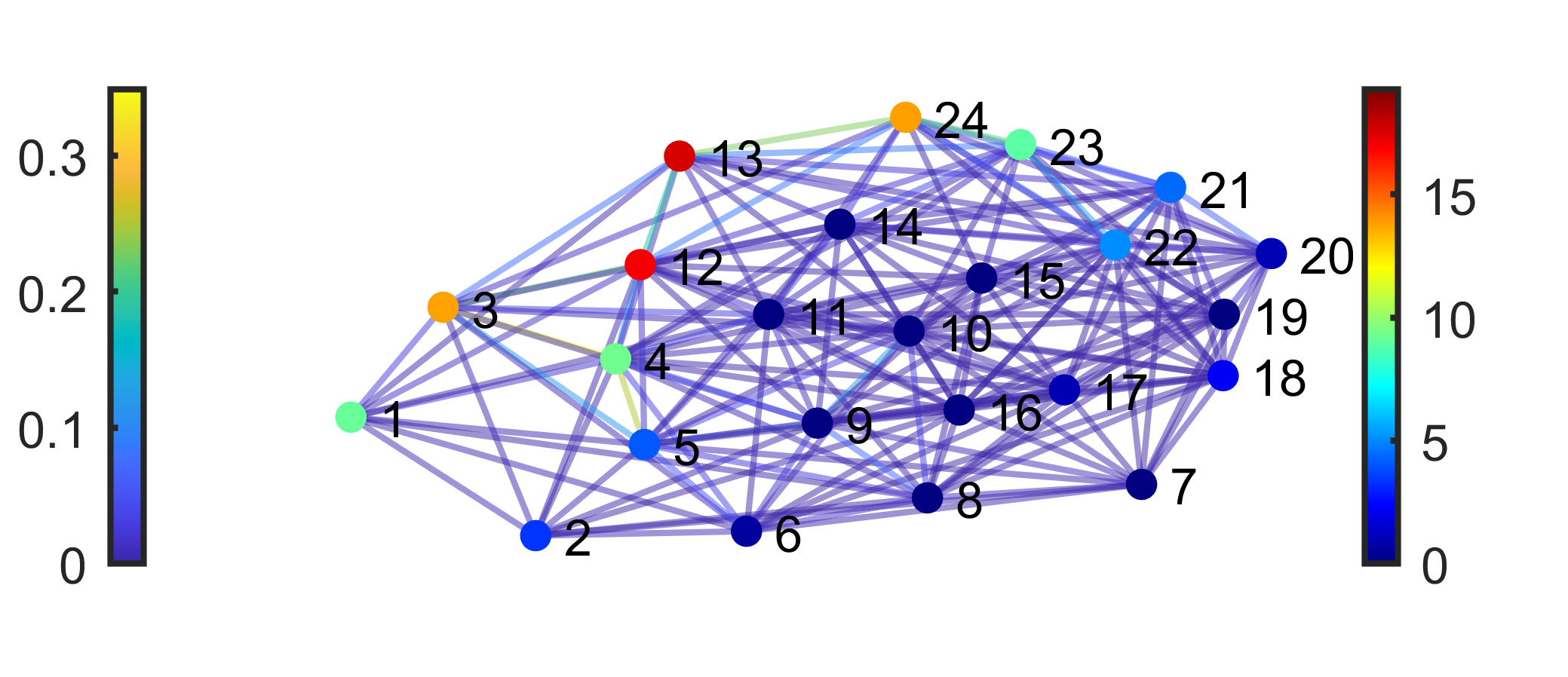}
		\vspace{-30pt} 
		\caption{$\mathcal{G}^{(p)}$ with $p = 3$}
		\label{fig_3_synthetic}
	\end{subfigure}
	\begin{subfigure}{0.45\textwidth}
		\centering
		\includegraphics[width=\textwidth]{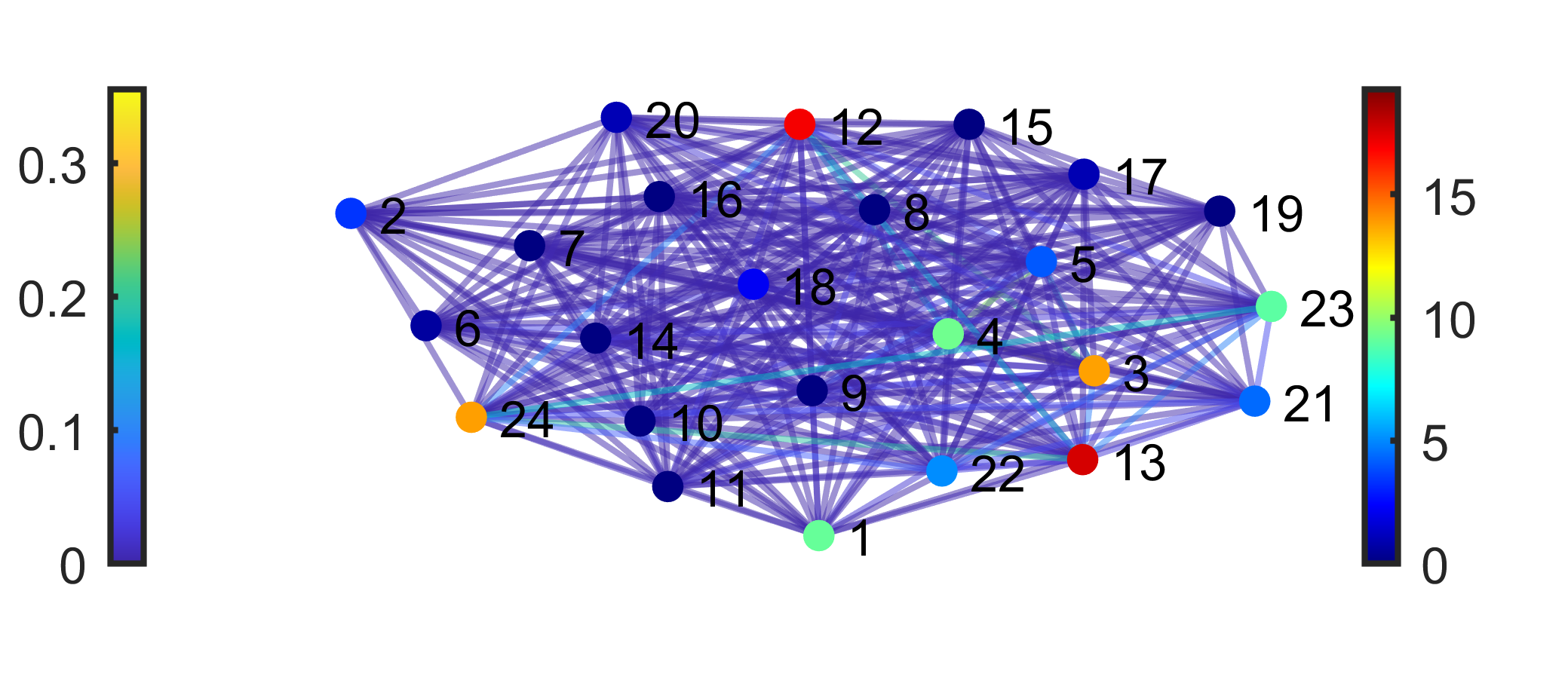}
		\vspace{-30pt}  
		\caption{$\mathcal{G}^{(p)}$ with $ p = 5$}
		\label{fig_5_synthetic}
	\end{subfigure}
	\caption{The original graph topology $\mathcal{G}$ and the topologies formed by directly connecting $p$-hop neighbors $\mathcal{G}^{(p)}$.  Left color bar: edges. Right color bar: nodes.}
	\label{fig_hops_135_t0}
\end{figure}

In this section, we formulate a data-driven approach to construct a dynamic topology from time-varying node and edge signals on a static initial graph. 
Dynamic Multi-hop captures node signal dynamics, forming dynamic graphs with time-varying edges that represent signal diffusion over time.
The illustration is conducted through an initial static graph $\mathcal{G}$ containing $N_n = 24$ nodes and $N_e = 38$ edges with its topology shown in Figure~\ref{fig_original_synthetic}. 

\subsection{P-Hop Node Signal Aggregation and Diffusion}
\label{sec_capture_node_diffusion}
The core of our Dynamic Multi-hop model is the diffusion (aggregation) of time-varying graph signals.
Considering the graph adjacency matrix, the matrix element ${a}_{ij}$ is non-zero only when there is an adjacent node $v_j$ to node $v_i$ that is 1 edge away. 
Similarly, the Laplacian matrix records non-zero element ${l}_{ij}$ when an edge exists between nodes$v_i$ and $v_j$, and non-zero diagonal entries to account for node degrees.
To illustrate the diffusion process, setting $P=1$, $\theta_1 = 1$, and $\theta_0 = 0$ in \eqref{eq_L_convolution}, setup leads to the following operation
\begin{equation}
	\boldsymbol{x}' = \mathbf{L}\boldsymbol{x} \in \mathbb{R}^{N_n}.
	\label{eq_L_agg}
\end{equation}
When \eqref{eq_L_agg} is viewed element-wise, the $i^{th}$ element represents the aggregation on nod $v_i$:
\begin{equation}
	{x}'_{i} = \sum^{N_n}_{j=1}{l}_{ij}x_{i},
	\label{eq_L_agg_element}
\end{equation}
where ${l}_{ij} = -w_{ij}$ is the element on the $i^{th}$ row and $j^{th}$ column of the Graph Laplacian matrix $\mathbf{L}$ and the value assignment is the negative of the weight of the corresponding edge. 
Equation \eqref{eq_L_agg} essentially repeats the element-wise operation shown in \eqref{eq_L_agg_element} for all the nodes: each summation effectively captures the local diffusion of signals across the 1-hop neighborhood along with self-aggregation.

Let us revisit \eqref{eq_L_agg} but now recursively apply it $P$ times:
\begin{equation}
	\mathbf{L}^P\boldsymbol{x}^{(0)} = \mathbf{L}^{P-1}(\mathbf{L}\boldsymbol{x}^{(0)}) = \mathbf{L}^{P-1}\boldsymbol{x}^{(1)} = \mathbf{L}^{P-2}\boldsymbol{x}^{(2)} = \dots ,
	\label{L_mult}
\end{equation}
where we denote the original node signal as $\boldsymbol{x} = \boldsymbol{x}^{(0)}$ and the diffused signal after $p$-hops as
\begin{equation}
	\boldsymbol{x}^{(p+1)}
	= \mathbf{L}\left[ \--\left(x_{i}^{(p)} - \sum^{N_{j}}_{j = 1} w_{ij}x_j^{(p)}\right)\-- \right]^T = \mathbf{L}\boldsymbol{x}^{(p)}.
	\label{L_mult_inside}
\end{equation}
The summation $\sum^{N_{j}}_{j = 1}$ essentially sums all the $N_j$ 1-hop adjacent nodes to $v_i$.
Traversing how the signals propagate \eqref{L_mult} and \eqref{L_mult_inside}, we see that for a node $v_{i}$, each of the neighborhood nodes $v_{j}$ got their signal components $x_{j}$ diffused to node $v_{i}$ $P$ times. 
This formulation will aggregate the data from the 1-hop neighbors of each node in such a manner that the aggregation weights are determined to minimize the difference between the 1-hop neighbors, which is, in turn, a diffusion operation as mentioned in \cite{stankovic_2019_vertex}. 
There are several choices of graph shift operators on undirected graphs, for example, the (weighted or unweighted) adjacency matrix, the (weighted or unweighted) graph Laplacian matrix (with no normalization), or the normalized graph Laplacian matrix. 
The difference between using the Laplacian matrix $\mathbf{L}$ and the adjacency matrix $\mathbf{A}$ is that the diagonal entries of the adjacency matrix are all zeros, meaning that the signal at the nodes themselves are not shifted. 
Following the 1-Hop logic shown previously, we can observe that the $p^{th}$ power of $\mathbf{L}$ in equation~\eqref{eq_L_agg} aggregates the simplicial signal from the $p$-hop neighbors.  
Take the graph Laplacian as a specific example, repeating \eqref{eq_L_agg} $p$ times we will be aggregating the weighted $p$-hop neighboring node signals and subtracting them from a weighted version of the signal at each node. 
Doing this $p$-hop aggregation allows algorithms to aggregate signals that are $p$-neighbors away.


Now, if we break down $\mathbf{L}^p$ again, we see that at several positions where there was once a zero in $\mathbf{L}$ becomes non-zero in $\mathbf{L}^p$. 
In Dynamic Multi-hop, we further take advantage of the non-zero entries by viewing them as latent edges representing multi-hop node diffusion.
Then, we can construct a series of new graphs by establishing direct edge connections from the newly appeared non-zero entries in $\mathbf{L}^p$.
Intuitively, suppose there is a non-zero entry in the $ij^{th}$ entry of $\mathbf{L}^p$ that was previously a zero $ij^{th}$ entry of $\mathbf{L}$. 
In this case, we connect an edge between the node $v_i$ and the node $v_j$.
After repeating this process for all nodes we denote the newly formed (static) graph topologies as $\mathcal{G}^{(p)}$ for each value of $p$. Their Laplacian matrices and adjacency matrices are denoted as $\mathbf{L}^{(p)}$ and $\mathbf{A}^{(p)}$ respectively. 
Mathematically, the above process of obtaining the adjacency matrix $\mathbf{A}^{(p)}$ of $\mathcal{G}^{(p)}$ is achieved by
\begin{equation}
	\mathbf{A}^{(p)}(\mathcal{G}^{(p)}) = \left(\mathbf{A}(\mathcal{G})\right)^p = \mathbf{A}^p,
	\label{eq_dense_adj}
\end{equation}
with $\mathcal{G}^{(1)}$ being the original graph $\mathcal{G}$.
A visual comparison of the topology formed by a weighted graph Laplacian $\mathbf{L}$ with $\mathbf{L}^3$ and $\mathbf{L}^5$ is shown in Figure~\ref{fig_hops_135_t0}. 

Each $\mathcal{G}^{(p)}$ forms denser graph topologies as we increase the number of hops $p$.
Viewing from a graph diffusion perspective, due to the fact that the edges between nodes are representations of the relationship and interaction of nodes, forming additional unweighted edges simply from expanding the computation of $\mathcal{G}^{(p)}$ by adding the number of hops is essentially adding irrelevant nodes into the local aggregation in \eqref{eq_L_agg}.
Increasing the number of hops will result in a graph over-smoothing scenario when the resulting aggregation of all the nodes in a graph will have similar or even the same value \cite{Chen_oversmoothing_2020}. 
This dense graph formulation often results in having less representation power compared to a sparse topology yet more representational graph  \cite{Ortega_graph_2018}. 
Thus, the obtained $\mathcal{G}^{(p)}\vert_{p = 2 \dots P}$ have to be pruned in order to maintain sparsity. 
Moreover, in the case when the signals nodes $\boldsymbol{x}[t]$ are dynamic, simply assigning time-invariant weights to the edges may not effectively capture node evolution and interaction patterns over time.
A more suitable approach would be to consider the weights between nodes as time-varying signals on the edges $\boldsymbol{w}[t] = [w_1[t] \dots w_{N_e}[t]]$.

\begin{figure*}
	\centering
	\begin{subfigure}{0.5\textwidth}
		\includegraphics[width=\textwidth]{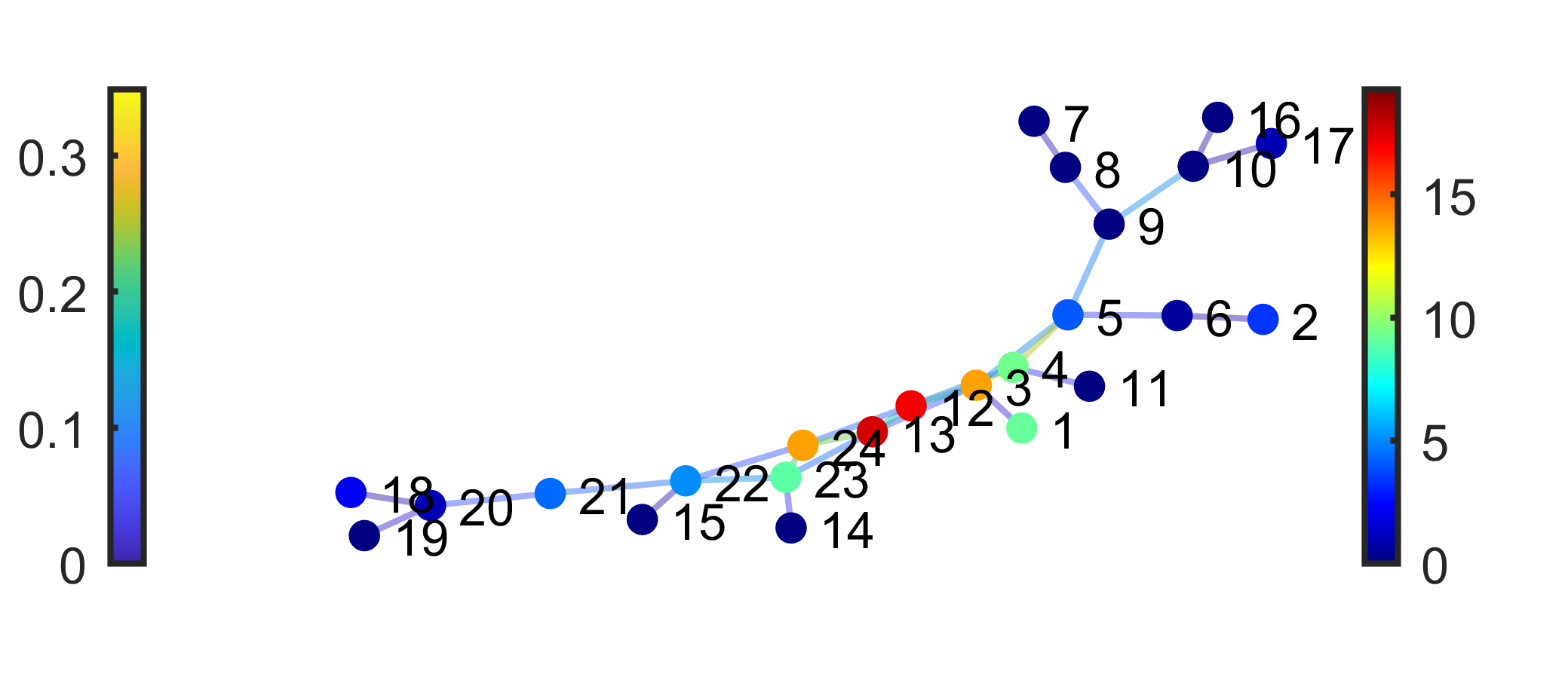}
		\vspace{-30pt}
		\caption*{$\bar{\mathcal{G}}^{(p)}[t]$ with $p = 3$ at $t = 1$.}
	\end{subfigure}
	\hspace{-10 pt}
	\begin{subfigure}{0.5\textwidth}
		\includegraphics[width=\textwidth]{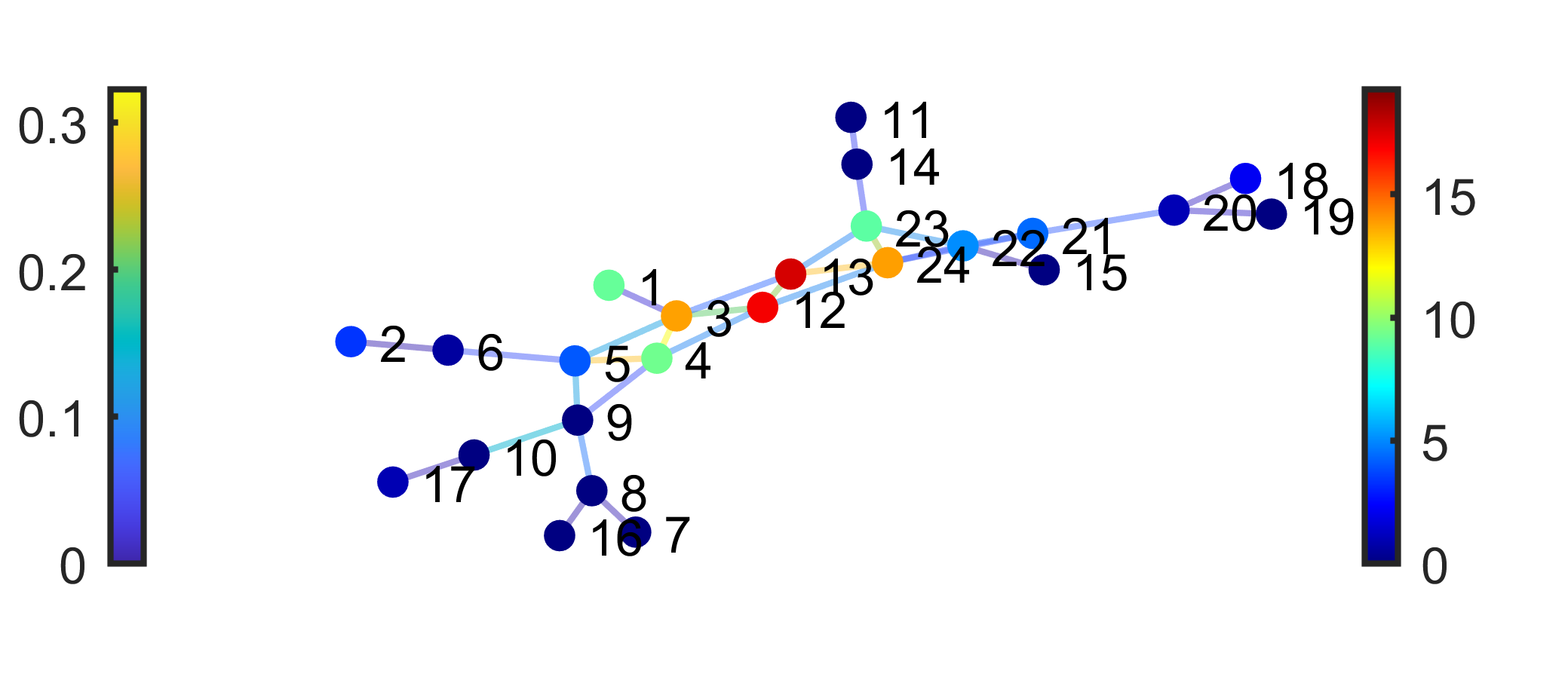}
		\vspace{-30pt}
		\caption*{$\bar{\mathcal{G}}^{(p)}[t]$ with $p = 3$ at $t = 30$.}
	\end{subfigure}
	
	\begin{subfigure}{0.5\textwidth}
		\includegraphics[width=\textwidth]{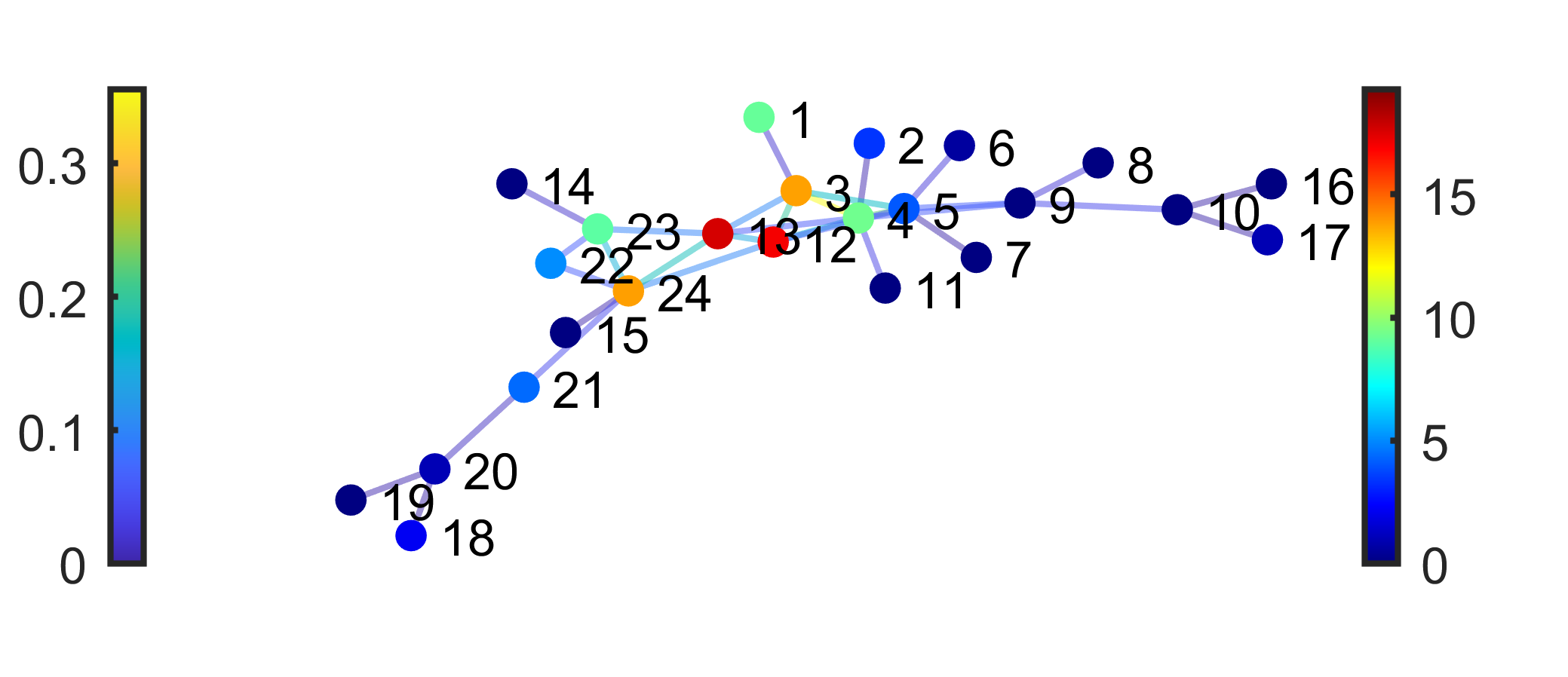}
		\vspace{-30pt}
		\caption*{$\bar{\mathcal{G}}^{(p)}[t]$ with $p = 6$ at $t = 1$.}
	\end{subfigure}
	\hspace{-10 pt}
	\begin{subfigure}{0.5\textwidth}
		\includegraphics[width=\textwidth]{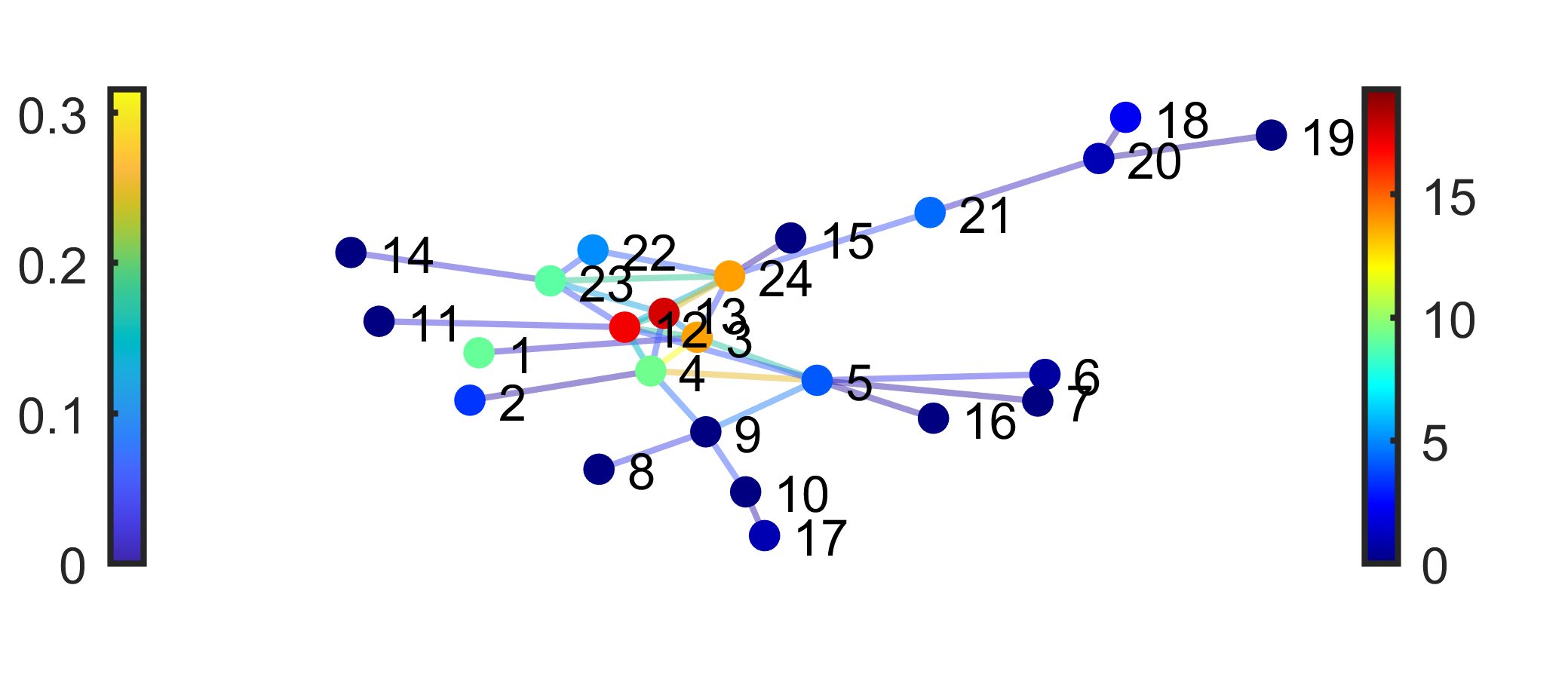}
		\vspace{-30pt}    
		\caption*{$\bar{\mathcal{G}}^{(p)}[t]$ with $p = 3$ at $t = 30$.}
	\end{subfigure}
	\caption{The topologies $\bar{\mathcal{G}}^{(p)}[t]$ formed at two time instances $t=0$ and $t=30$ by directly connecting $p$-hop neighbors then pruning the edges with time-varying weights in $\mathcal{G}^{(p)}$.  Left color bar: edges. Right color bar: nodes.}
	\label{fig_hop_pruned}
\end{figure*}

\subsection{From Time-Varying Edge Signals To Dynamic Multi-hop}
Let us discuss how considering edge weights as time-varying signals on the edges could change the story and overcome the limitations mentioned in the previous subsection. 
Inspired by SGMs, in Dynamic Multi-hop we apply a pruning-based approach in which the topologies are formed by calculating statistical measures among data and then taking a threshold to eliminate edges between weakly correlated nodes \cite{zhang2015graph}. 
We assume that the time-varying weights $\boldsymbol{w}^{(p)}[t]$ of the underlying graph topologies $\mathcal{G}^{(p)}$ also reflects a statistical measure of the node interactions and are non-negative to maintain the properties of the graph Laplacian matrix.
For example, in the GGM, the edges can be formed using the precision matrix, which is the inverse of the covariance matrix of a multivariate Gaussian distribution \cite{zhang2015graph}. 

In order to represent edge weights as signals on the graph edges, we will be relying on techniques from TSP \cite{Barbarossa_2020}. 
For a given edge signal $\boldsymbol{w}$, one can process it similar to the signals on the nodes shown in \eqref{eq_L_convolution} by \cite{Kadambari_2022_Distributed}
\begin{equation}
	\mathcal{C}(\mathbf{L}_1, \mathcal{F}) \boldsymbol{w} = \mathbf{U}_1 \mathbf{\Sigma}_{\mathcal{F}, 1}\mathbf{U}_1^T\boldsymbol{w} \approx \sum_{p=0}^{P} \theta_p \mathbf{L}_1^p \boldsymbol{w}. 
	\label{eq_L_convolution_edge}
\end{equation}
In \eqref{eq_L_convolution_edge}, $\mathbf{U}_1$ is the eigenvector matrix of the Hodge $1$ Laplacian $\mathbf{L}_1\in \mathbb{R}^{N_e \times N_e}$ (also known as the graph Helmholtzian) and $\mathbf{\Sigma}_{\mathcal{F}, e}$ is a spectral filter on the edges.
Using TSP notation, in the presence of node and edge signal but with the absence of triangle signal, $\mathbf{L}_1$ can be defined as
\begin{equation}
	\mathbf{L}_1 = \mathbf{B}_1^T\mathbf{B}_1,
	\label{eq_L1_def_B}
\end{equation}
where $\mathbf{B}_1 \in \mathbb{R}^{N_n \times N_e}$ is the node-to-edge incidence matrix defined by
\begin{equation}
	\mathbf{L} = \mathbf{B}_1\mathbf{B}_1^T = \text{sign}(\mathbf{B}_1) \text{diag}(\boldsymbol{w}) \text{sign}(\mathbf{B}_1^T).
	\label{eq_L_def_B}
\end{equation}

When the edge weights $\boldsymbol{w}[t]$ are time-varying, the graph Laplacian matrices become time-varying ${\mathbf{L}}[t]$ as well:
\begin{equation}
	\mathbf{L}[t] = \mathbf{B}_1[t]\mathbf{B}_1[t]^T = \text{sign}(\mathbf{B}_1) \text{diag}(\boldsymbol{w}[t]) \text{sign}(\mathbf{B}_1^T).
\end{equation} 
We denote the graph with time-varying weights $\boldsymbol{w}[t]$ as $\mathcal{G}[t]$.
If we revisit the $P$-hop node signal diffusion \eqref{L_mult} and break it down using the definitions in \eqref{eq_L1_def_B} and \eqref{eq_L_def_B}, we can observe that the $P$-hop node signal diffusion \eqref{L_mult} implicitly expresses the $(P-1)$-hop edge diffusion:
\begin{equation}
	\begin{split}
		\mathbf{L}[t]^P \boldsymbol{x}[t] &= (\mathbf{B}_1[t]\mathbf{B}_1[t]^T)^P \boldsymbol{x}[t] \\&= \mathbf{B}_1[t] (\mathbf{B}_1[t]^T\mathbf{B}_1[t])^{(P-1) }\mathbf{B}_1[t]^T \boldsymbol{x}[t] \\ &= \mathbf{B}_1[t] \mathbf{L}_1[t]^{(P-1)} \mathbf{B}_1[t]^T \boldsymbol{x}[t].
	\end{split}
\end{equation}
Following the logic from Section~\ref{sec_capture_node_diffusion}, realizing that for $p>1$, we can again represent them as new edges between the nodes to form new graphs using the newly appeared non-zero entries.
A series of new graphs, denoted as $\mathcal{G}^{(p)}[t]$ for $p = 2, \dots, P$, is constructed for each hop $p$ at each time instance $t$. 
Each of $\mathcal{G}^{(p)[t]}$ represents the node $p$-hop diffusion to reflect the node diffusion dynamics over time.
The newly formed connections in new graphs $\mathcal{G}^{(p)[t]}\vert_{p = 2 \dots P}$ are realizations of diffusion trajectories into edges, representing the far and latent interaction between node $v_i$ and node $v_j$ that was previously not captured by the compared to original topology $\mathcal{G}[t]$.
In $\mathcal{G}^{(p)}[t]$, an edge is placed between nodes $v_i$ and $v_j$ for every non-zero $ij^{\text{th}}$ entry in $\mathbf{L}[t]^p$ that was previously zero in $\mathbf{L}[t]$.
Each newly placed edge $\mathcal{G}^{(p)}[t]$ represents there is a hidden interaction between node $v_i$ and node $v_j$ obtained from $p$-hop diffusion that was previously not represented in the original graph $\mathcal{G}$.
In other words, $\mathcal{G}^{(p)}[t]$ captures the node interactions due to $p$-hop diffusion at time $t$.
However,  $\mathcal{G}^{(p)}[t]$ is still dense, whereas sparse graphs are more desirable in GSP.

The key to achieving a sparse yet representative graph in Dynamic Multi-hop is to construct a topology with only a few additional edges compared to the original graph $\mathcal{G}$. 
Each new edge represents critical connections that capture the node diffusion dynamics at time $t$.
In SGM, a typical method to construct a graph topology is pruning the correlation matrix or the precision matrix using a threshold \cite{zhang2015graph}. 
In Dynamic Multi-hop, we take a similar approach to prune the dense graphs $\mathcal{G}^{(p)}[t]$ for $p = 2, \dots, P$, keeping only the edges that have signals reflecting strong node interactions at time $t$.
A pruning problem can be formulated as follows, aiming to minimize the number of edges by keeping only the edges for which the pruning metric $S(e_i[t])$ exceeds a given threshold $\tau$:
\begin{equation}
	\min_{\bar{\mathcal{E}}^{(p)}[t]} \sum_{e_i \in \mathcal{E}^{(p)}}^{N_e} \mathbf{1}(S(e_i[t]) > \tau)\Bigg\vert_{p = 1 \dots P}.
	\label{eq_pruning_cost}
\end{equation}
In \eqref{eq_pruning_cost}, the set $\bar{\mathcal{E}}^{(p)}$ represents the edge set for the topology $\bar{\mathcal{G}}^{(p)}$ and the pruning criterion for each edge is denoted by the function $S(e_i[t])$.
Here in \eqref{eq_pruning_cost}, the pruning criterion is represented as a generalized term  $S(e_i[t])$, aiming to be a user-defined specification following the requirements of the downstream task.  
Following \eqref{eq_pruning_cost}, we obtain the pruned version of $\mathcal{G}^{(p)}[t]$ as $\bar{\mathcal{G}}^{(p)}[t]$ by removing edges $e_i$ from the edge set $\mathcal{E}^{(p)}[t]$ where the pruning metric $S(e_i[t])$ is less than or equal to the threshold $\tau$:
\begin{equation}
	\bar{\mathcal{G}}^{(p)}[t] = \mathcal{G}^{(p)}[t] \setminus \{ e_i \in \mathcal{E}^{(p)}[t] \mid S(e_i[t]) \leq \tau \}.
\end{equation}

\begin{figure*}
	\begin{subfigure}{0.5\textwidth}
		\includegraphics[width=\textwidth]{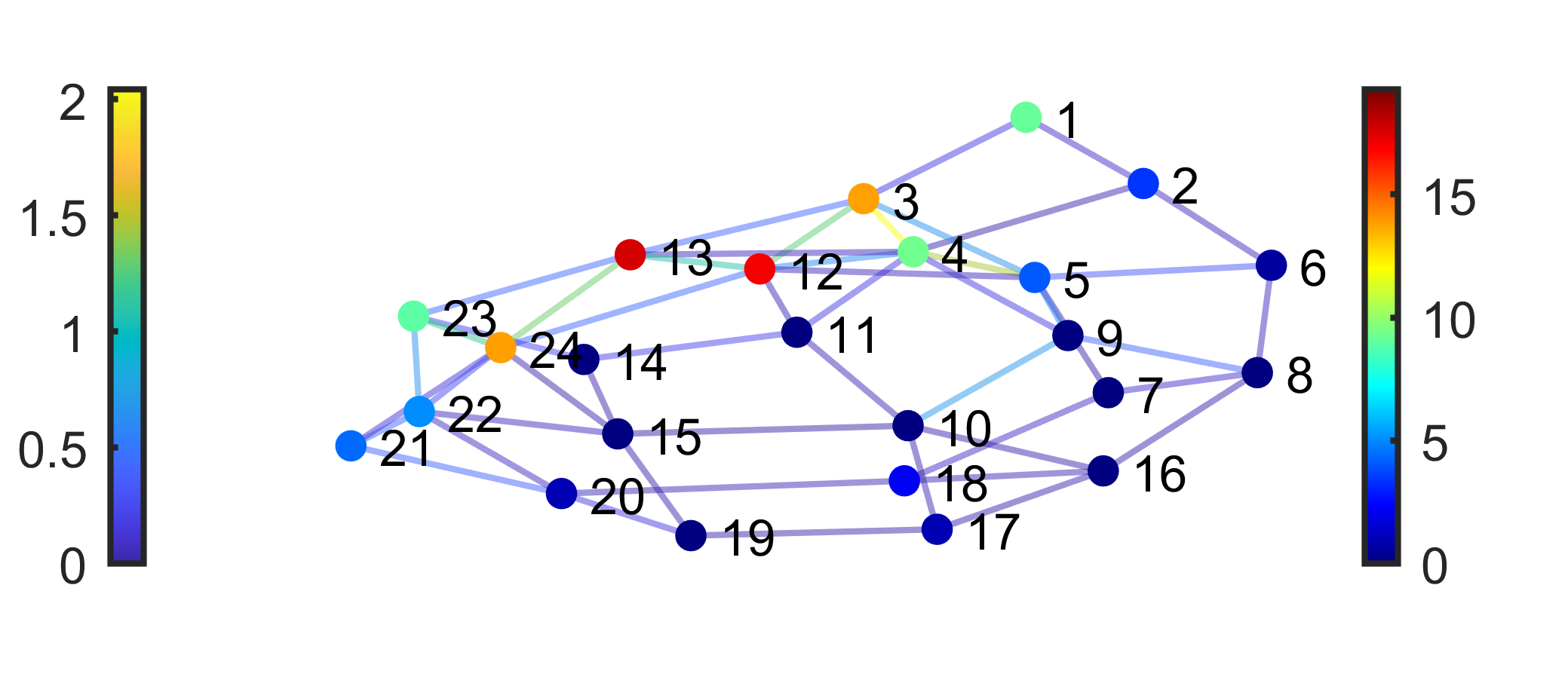}
		\vspace{-30pt}         
		\caption*{$\bar{\mathcal{G}}[t]$ at $t = 1$.}
	\end{subfigure}
	\hspace{-10 pt}
	\begin{subfigure}{0.5\textwidth}
		\includegraphics[width=\textwidth]{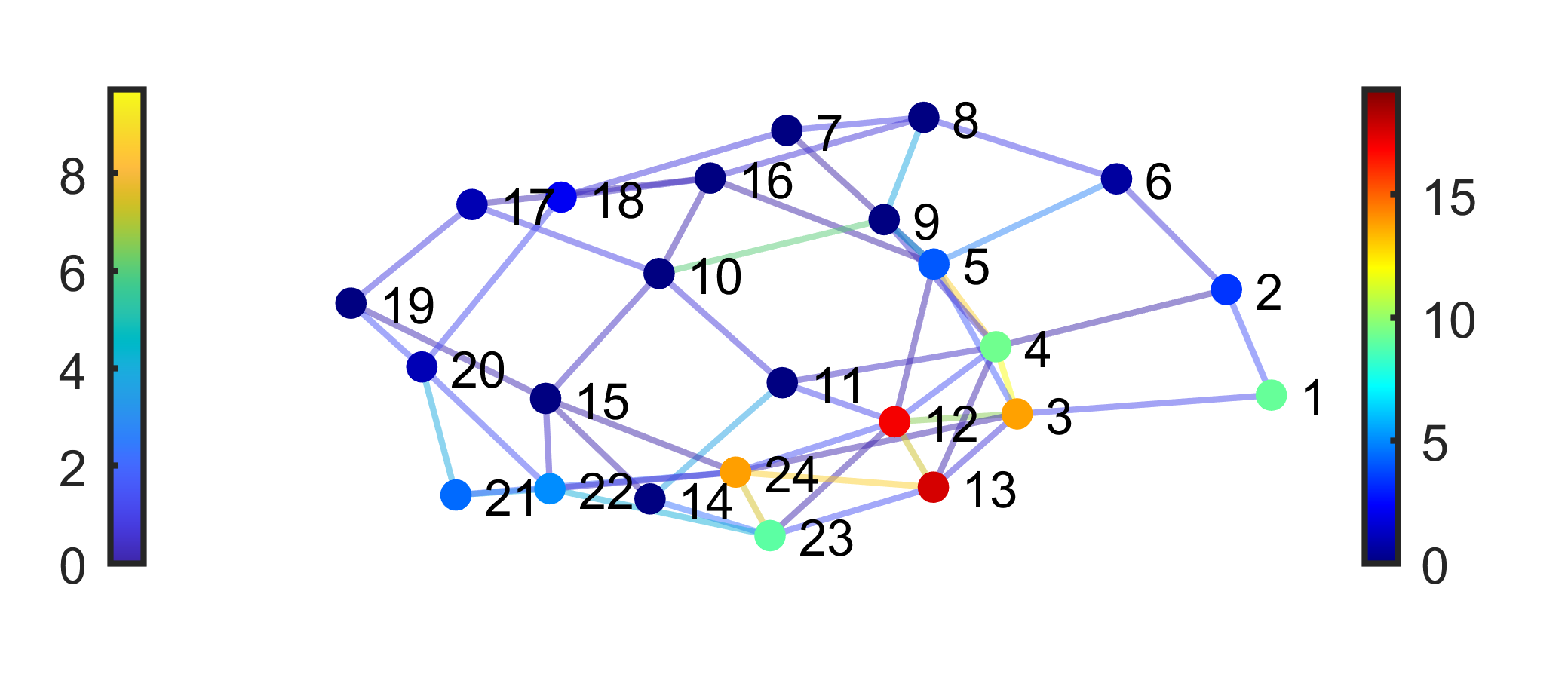}
		\vspace{-30pt}         
		\caption*{$\bar{\mathcal{G}}[t]$ at $t = 30$.}
	\end{subfigure}
	\caption{The topologies $\bar{\mathcal{G}}[t]$ formed at two time instances $t=0$ and $t=30$ by directly merging all $\bar{\mathcal{G}}^{(p)}$s for $p = 1\dots6$. Left color bar: edges. Right color bar: nodes.}
	\label{fig_merged_hops}
\end{figure*}

Given that the weights $\boldsymbol{w}[t]$ are time-varying, different edges can be eliminated from $\mathcal{G}^{(p)}[t]$ at different time instances if the threshold $\tau$ designed properly.
Each resulting pruned graph $\bar{\mathcal{G}}^{(p)}[t]$ is a sparse graph that has only a few edges representing the important $p$-hop node diffusion that have higher values than the threshold $\tau$ at time $t$. 
A demonstration of forming dynamic topologies $\bar{\mathcal{G}}^{(p)}[t]$ from pruning dense graph $\mathcal{G}^{(p)}[t]$ is shown in Figure~\ref{fig_hop_pruned}.
As we mentioned earlier, each $\bar{\mathcal{G}}^{(p)}[t]$ represents the important $p$-hop node diffusion that have higher values than the threshold $\tau$, while maintaining a sparser topology relative to $\mathcal{G}^{(p)}[t]$. 
We can represent all the important node diffusion for $p = 1, \dots, P$ by merging all the pruned topologies $\bar{\mathcal{G}}^{(2)}, \dots, \bar{\mathcal{G}}^{(P)}$ and stacking them on top of the original graph. In this way, Dynamic Multi-hop forms a sparse graph $\bar{\mathcal{G}}[t]$ at time $t$, which incorporates the latent interactions of 1-hop to $P$-hop node interactions:
\begin{equation}
	\bar{\mathcal{G}}[t] = \mathcal{G}[t] \cup \bigcup_{p=2}^{P} \bar{\mathcal{G}}^{(p)}[t].
	\label{eq_merge}
\end{equation}
An illustration of the merged topology that resulted in a dynamic graph is shown in Figure~\ref{fig_merged_hops}. 

The rationale behind the pruning and merging is that compared to the original graph topology $\mathcal{G}$, each $\bar{\mathcal{G}}[t]$ at different time instances should have a few new edges that represent the hidden node interactions that were previously absent in $\mathcal{G}$.
It should be emphasized that we can get a dynamic graph through pruning $\mathcal{G}^{(p)}$ only when the edge signals are time-varying.
The reason behind this is that pruning the edge on a graph with static weights always results in cutting the same edges at different time instances. 
However, if we prune edges based on the evolving numerical value of the time-varying edge signals $\boldsymbol{w}[t]$, then we are likely to cut different edges at different time instances. 
The simultaneous utilization of time-varyingness, SGM, graph pruning, and TSP is the unique characteristic of Dynamic Multi-hop. 
Lastly, we would like to point out that even if we already started with a series of dynamic graphs, we can still apply the Dynamic Multi-hop model to reveal several additional edges that were previously absent from the given dynamic graphs to enhance the representation power.
The Dynamic Multi-hop provides an effective foundation that enhances the modeling of complex spatiotemporal dependencies and allows the capturing of time-varying graph signal interactions.
In the next section, we extend Dynamic Multi-hop to incorporate adaptive estimations, enabling more precise tracking of time-varying signals.

\section{Online Estimation of Node Signal with Dynamic Multi-hop}
\label{sec_adaptive}
Building on the dynamic topology formulation, we now extend the Dynamic Multi-hop model to facilitate online estimation of time-varying node signals, leveraging techniques from GSP to enhance adaptability.
Let us begin with a universal data model to express the time-varying characteristics of the node signal:
\begin{equation}
	\boldsymbol{x}[t+1]-\boldsymbol{x}[t] = \Delta[t],
	\label{eq_data_model}
\end{equation}
where $\Delta[t]$ is the amount of change that will lead the current graph signal $\boldsymbol{x}[t]$ to the next step graph signal $ \boldsymbol{x}[t+1]$. 
Assuming that we are using an adaptive algorithm similar to the previously proposed GLMS \cite{bib_LMS}, GLMP \cite{nguyen2020_LMP}, or GSD \cite{yan_2023_sign}, the cost function of the Dynamic Multi-hop estimation is
\begin{equation}
	\min_{\hat{\boldsymbol{x}}[t]} J(\hat{\boldsymbol{x}}[t]),
	\label{eq_cost_adaptive}
\end{equation}
where $J$ is a convex function. 
Following the convention in GSP, suppose we have a noisy observation $\boldsymbol{y}[t] = \mathbf{M}[t](\boldsymbol{x}[t]+\boldsymbol{\eta}[t])$, where $\boldsymbol{x}[t]$ is the ground truth signal, $\mathbf{M}[t]$ is the observation mask to mimic partial observation, and $\boldsymbol{\eta}[t]$ is the noise in observation \cite{Spelta_2020_NLMS}. 
If the goal is to obtain online estimations $\hat{\mathbf{x}[t]}$ of the signal $\boldsymbol{x}[t]$ in the minimum mean square error (MSE) sense using the dynamic topologies formed by the Dynamic Multi-hop.
Then, the cost function $J(\hat{\boldsymbol{x}[t]})$ becomes
\begin{equation}
	J(\hat{\boldsymbol{x}}[t]) = \mathbb{E}\left\|\boldsymbol{y}[t]-\mathbf{M}\mathcal{C}(\bar{\mathbf{L}}[t], \mathcal{F})\hat{\boldsymbol{x}}[t]\right\|_2^2,
\end{equation}
where $\bar{\mathbf{L}}[t]$ is the graph Laplacian of $\bar{\mathcal{G}}[t]$ and \begin{equation}
	\mathcal{C}(\bar{\mathbf{L}}[t], \mathcal{F}) = \bar{\mathbf{U}}[t]\mathbf{\Sigma}_\mathcal{F}\bar{\mathbf{U}}^T[t]\boldsymbol{x}[t] \approx \sum_{p=0}^{P} \theta_p \bar{\mathbf{L}}^p[t] \boldsymbol{x}[t].
	\label{eq_adaptive_filter}
\end{equation}
The solution to \eqref{eq_cost_adaptive} can be obtained by taking the gradient and utilizing the bandlimited property from the filter $\mathbf{\Sigma}_\mathcal{F}$: 
\begin{equation}
	\frac{\partial J(\hat{\boldsymbol{x}}[t])}{\partial\hat{\boldsymbol{x}}[t]} =-2\mathcal{C}(\bar{\mathbf{L}}[t], \mathcal{F})\mathbf{M}(\boldsymbol{y}[t]-\hat{\boldsymbol{x}}[t]).
	\label{eq_optimization_result}
\end{equation}
We can then plug ${\mathcal{C}}(\bar{\mathbf{L}}[t], \mathcal{F})$ into the optimization result \ref{eq_optimization_result} to form an adaptive filter graph signals estimation similar to the data model \eqref{eq_data_model}:
\begin{equation}
	\begin{split}
		\hat{\boldsymbol{x}}[t+1] = \hat{\boldsymbol{x}}[t]+ \mu {\mathcal{C}}(\bar{\mathbf{L}}[t], \mathcal{F})\mathbf{M}(\boldsymbol{y}[t]-\hat{\boldsymbol{x}}[t]).
		\label{eq_update}
	\end{split} 
\end{equation}
where $\mu$ is a step size parameter that controls the update magnitude as seen in classical adaptive filters \cite{Diniz_2007_adaptive_filtering}.

The adaptive graph signal estimation algorithm in \eqref{eq_update} integrates the Dynamic Multi-hop framework with principles from classical adaptive filtering, building on the LMS approach.
At each time instance $t$, the Dynamic Multi-hop estimation in \eqref{eq_update} utilizes the updated graph topology $\bar{\mathcal{G}}[t]$, which is a graph that reflects the recent diffusion interactions among the nodes at $t$, and makes online estimations in the direction that minimizes the MSE. 
Dynamic Multi-hop estimation stands out from previous GSP methods by incorporating the dynamic changes in node connections, inferred from the diffusion of time-varying node signals represented by the time-varying edge signals, revealing the latent interactions among the nodes that were not represented in static topology. 
This dynamic representation is more effective than static counterparts, as it adapts to the evolving nature of graph signals, leading to more accurate online graph signal estimation.
Notably, unlike classical adaptive filtering that updates the filters at each time step, the Dynamic Multi-hop estimation in \eqref{eq_update} uses a predefined filter as seen in GSP, avoiding the computations and instabilities of frequent filter updates \cite{yan_2022_sign}.
Previous adaptive-filter-based algorithms on graphs are proven to converge in the Mean Squared sense \cite{bib_LMS, nguyen2020_LMP, yan_2022_sign}.
Following similar as found in GSP methods, the Dynamic Multi-hop estimation can be shown that convergence can be achieved under the condition of $ 0 < \mu < \frac{2}{\lambda_\text{max}}$, where $\lambda_\text{max}$ is the maximum eigenvalue of $(\bar{\mathbf{U}}[t]\mathbf{\Sigma}_\mathcal{F})^T\mathbf{M}(\bar{\mathbf{U}}[t]\mathbf{\Sigma}_\mathcal{F})$. 
The procedure of using Dynamic Multi-hop in an adaptive filter to conduct an online estimation of graph signal is summarized in Algorithm~\ref{alg_dynamic_multi_hop_adaptive}. 

In practice, we can initialize parameter $\mu$ and filter $\mathbf{\Sigma}_\mathcal{F}$ of the Dynamic Multi-hop estimation using historical data or directly from the training set.
Subsequently, the Dynamic Multi-hop estimation can be deployed to test data or unknown data. 
In the deployment of Dynamic Multi-hop estimation, due to the different behavior of the graph signals and edge weights, data preprocessing techniques such as scaling, shifting, or normalization should be taken when necessary. 
For example, the edge weights or the Laplacian matrix could be normalized to prevent the $p^{th}$ power of the Laplacian matrix from exploding.
\begin{algorithm}
	\caption{Dynamic Multi-hop Estimation of Time-varying Graph Signal}\label{alg_dynamic_multi_hop_adaptive}
	\begin{algorithmic}[1]
		\State {\bf{Given}} static graph $\mathcal{G}$, filter $\mathbf{\Sigma}_\mathcal{F}$, and $P$
		\State Define pruning criterion $S(e_i)$ in \eqref{eq_pruning_cost}
		\While{there are new node signal observation $\boldsymbol{y}[t]$}
		\State Obtain $\boldsymbol{w}[t]$ or redefine $\boldsymbol{w}[t]$ if needed
		\State Obtain dense graphs $\mathcal{G}^{(p)}[t]\vert_{p = 1 \dots P}$ using \eqref{eq_dense_adj} and $\hat{\boldsymbol{x}}[t]$
		\State Prune $\mathcal{G}^{(p)}[t]\vert_{p = 1 \dots P}$ by \eqref{eq_pruning_cost} to produce $\bar{\mathcal{G}}^{(p)}[t]\vert_{p = 1 \dots P}$
		\State Get $\bar{\mathcal{G}}[t]$ to merge $\bar{\mathcal{G}}^{(p)}[t]\vert_{p = 1 \dots P}$ using \eqref{eq_merge}
		\State Define filtering operation by feeding  $\bar{\mathcal{G}}[t]$ into \eqref{eq_adaptive_filter}  
		\State Predict $\hat{\boldsymbol{x}}[t]$ using $\boldsymbol{y}[t]$ and \eqref{eq_update}
		\EndWhile
	\end{algorithmic}
\end{algorithm}

\begin{figure*}[h]
	\centering
	\begin{subfigure}{0.45\textwidth}
		\centering
		\includegraphics[trim= 0 30 0 30,clip,width=\textwidth]{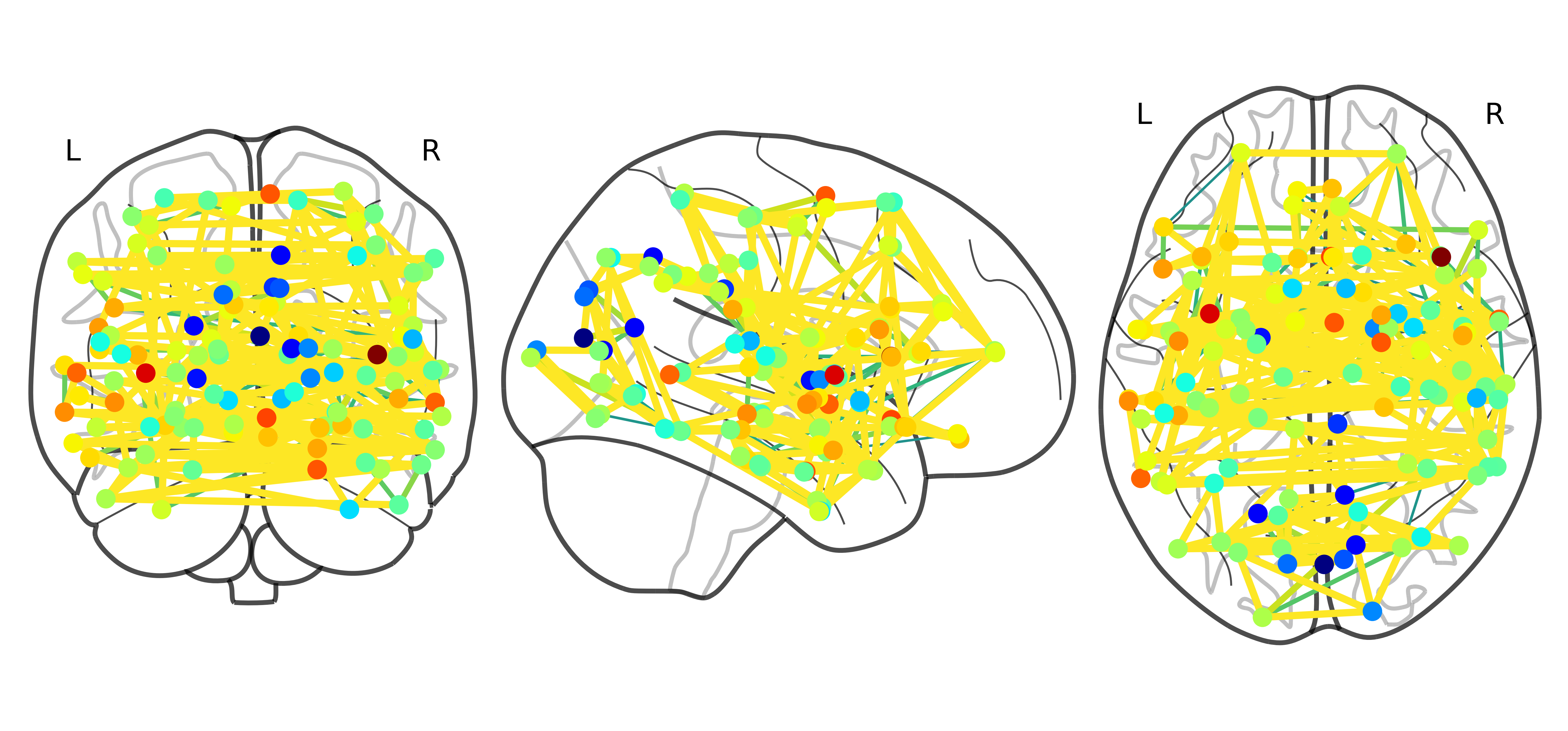}
		\caption*{Normal subject, $t = 52$.}
	\end{subfigure}
	\begin{subfigure}{0.45\textwidth}
		\centering
		\includegraphics[trim= 0 30 0 30,clip,width=\textwidth]{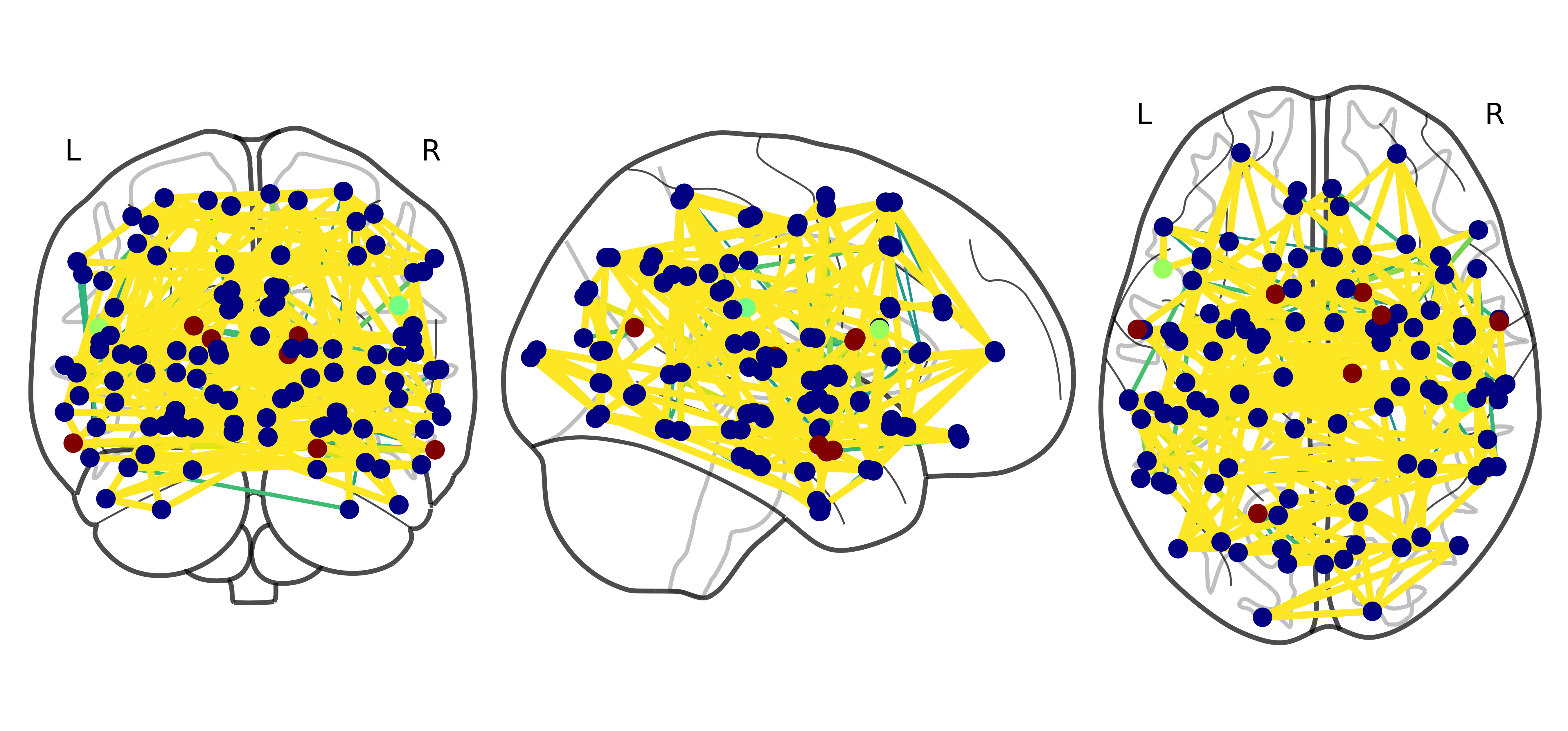}
		\caption*{Autistic subject, $t = 52$.}
	\end{subfigure}
	\begin{subfigure}{0.45\textwidth}
		\centering
		\includegraphics[trim= 0 30 0 30,clip,width=\textwidth]{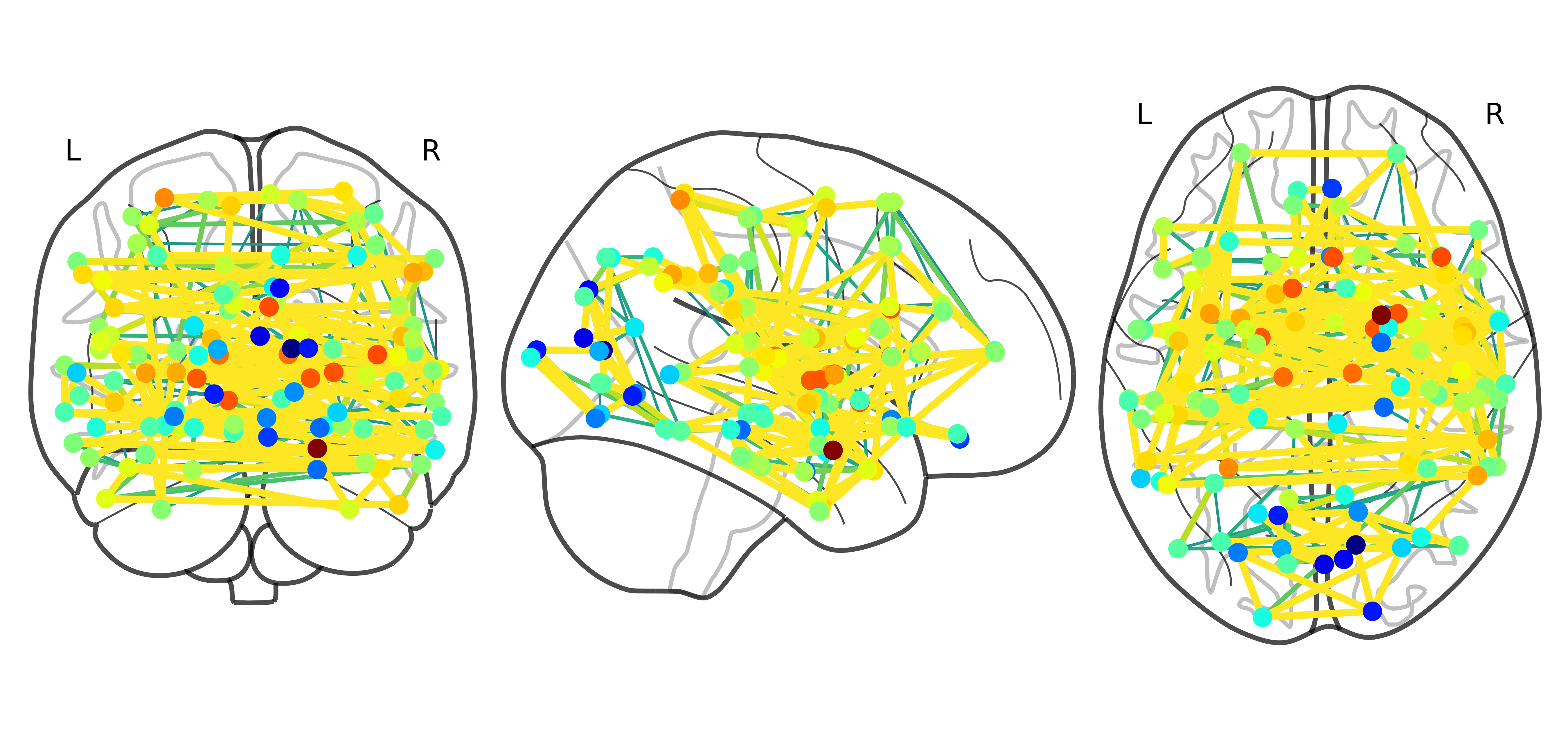}
		\caption*{Normal subject, $t = 60$.}
	\end{subfigure}
	\begin{subfigure}{0.45\textwidth}
		\centering
		\includegraphics[trim= 0 30 0 30,clip,width=\textwidth]{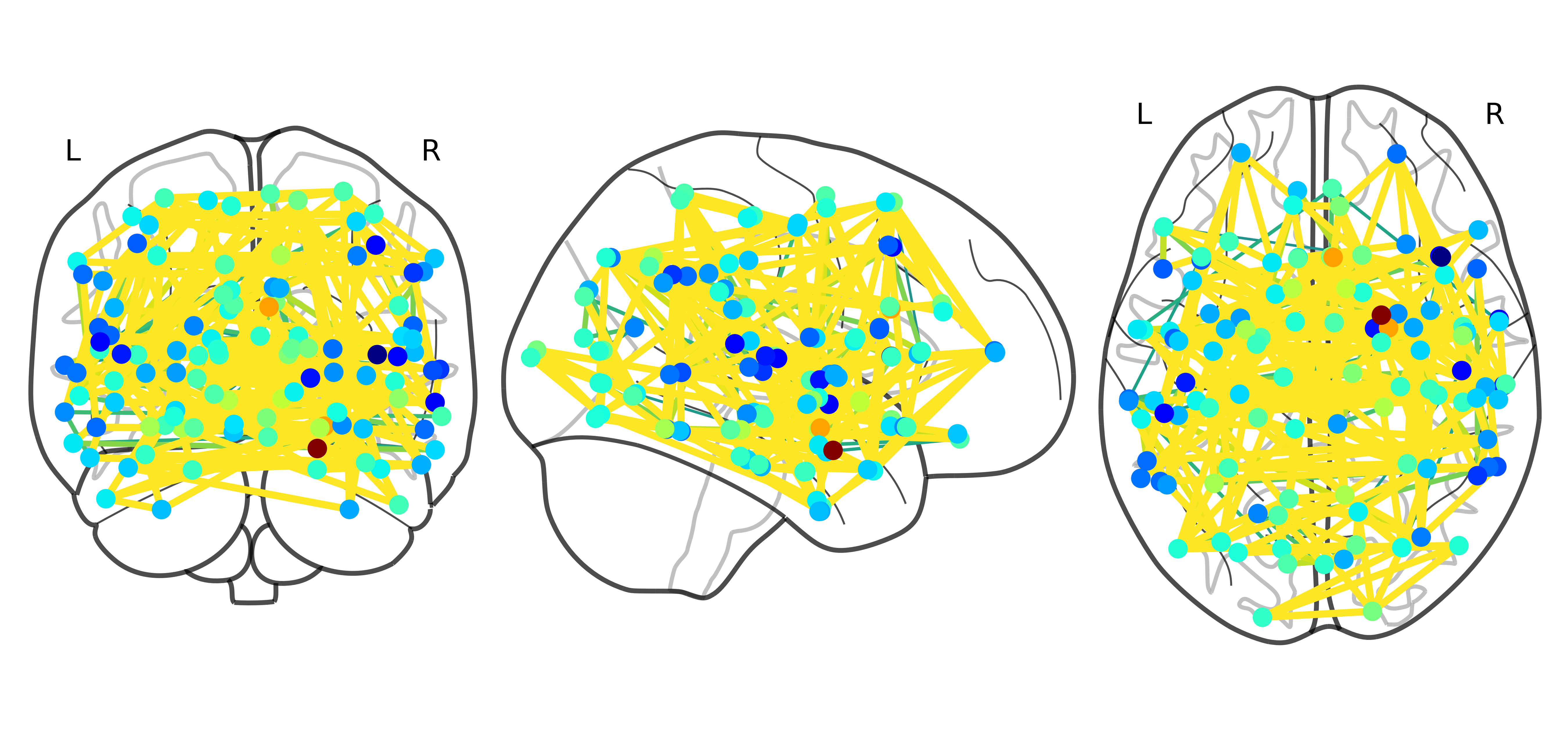}
		\caption*{Autistic subject, $t = 60$.}
	\end{subfigure}
	\caption{Dynamic brain networks $\bar{\mathcal{G}}[t]$ formed for both normal and autistic subjects by Dynamic Multi-hop.}
	\label{fig_brain_prune}
\end{figure*}

\begin{figure*}[h]
	\centering
	\begin{subfigure}{0.45\textwidth}
		\centering
		\includegraphics[trim= 0 30 0 30,clip,width=\textwidth]{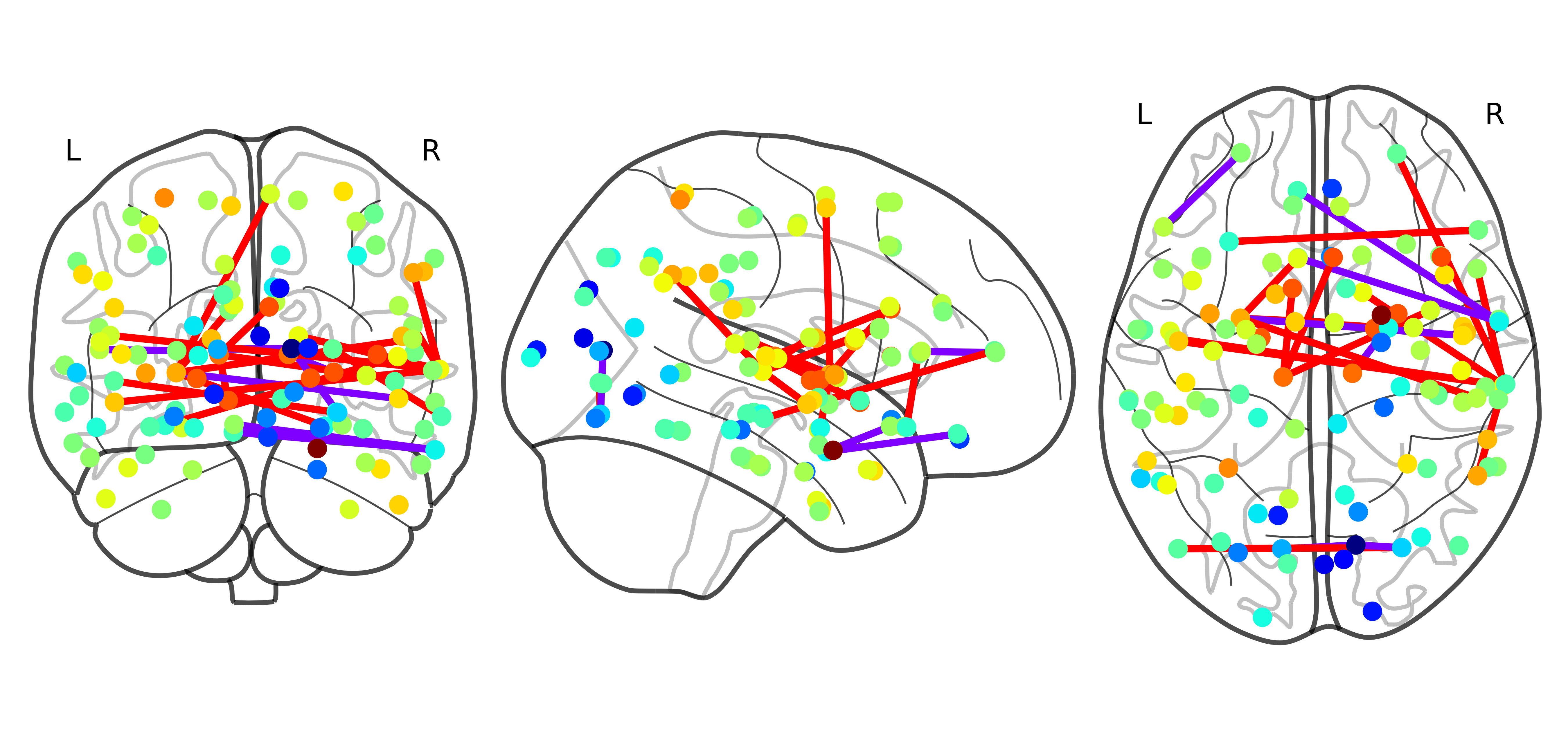}
		\caption*{Normal Subject}
	\end{subfigure}
	\begin{subfigure}{0.45\textwidth}
		\centering
		\includegraphics[trim= 0 30 0 30,clip,width=\textwidth]{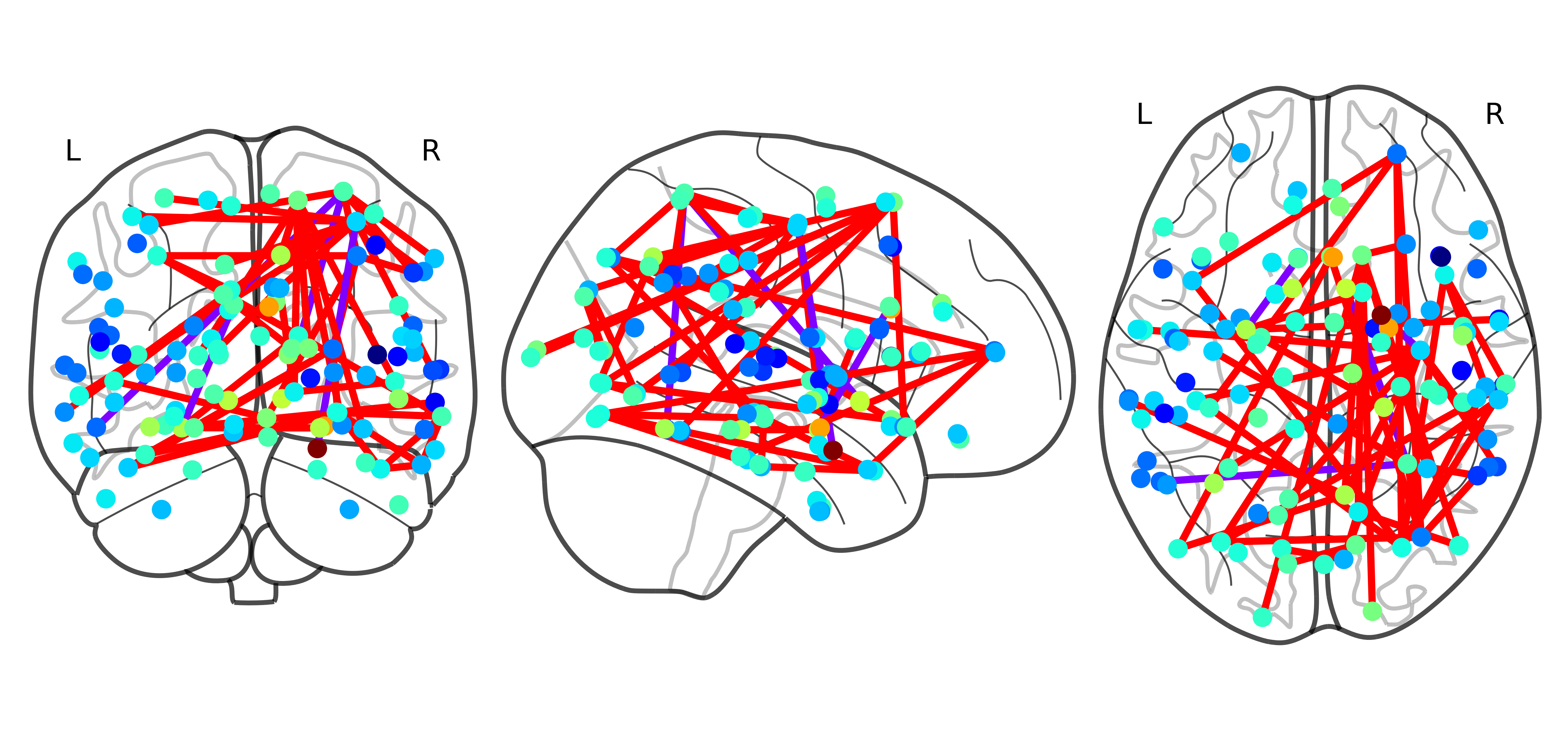}
		\caption*{Autistic subject}
	\end{subfigure}
	\caption{Comparison between the topology at $t = 52$ and $t = 60$ mapped on the brain image. The purple edges have appeared at $t = 52$ but not in $t = 60$ and the red edges edges have appeared at $t = 60$ but not in $t = 52$.}
	\label{fig_change}
\end{figure*}

\begin{figure*}[h]
	\centering
	\begin{subfigure}{0.32\textwidth}
		\centering
		\includegraphics[trim= 180 270 180 270,clip, width=\textwidth]{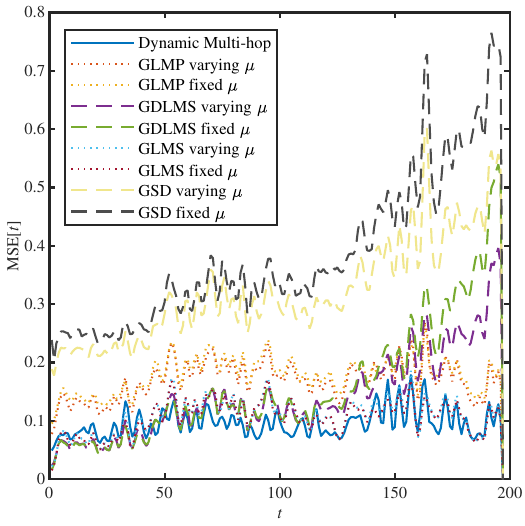}
		\vspace{-20 pt}
		\caption*{Nornal Subject, SNR = $3$.}
	\end{subfigure}
	\begin{subfigure}{0.32\textwidth}
		\centering
		\includegraphics[trim= 180 270 180 270,clip, width=\textwidth]{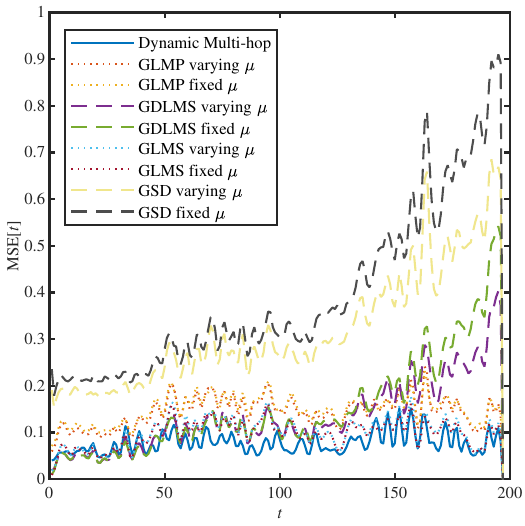}
		\vspace{-20 pt}
		\caption*{Normal  Subject, SNR = $5$.}
	\end{subfigure}
	\begin{subfigure}{0.32\textwidth}
		\centering
		\includegraphics[trim= 180 270 180 270,clip, width=\textwidth]{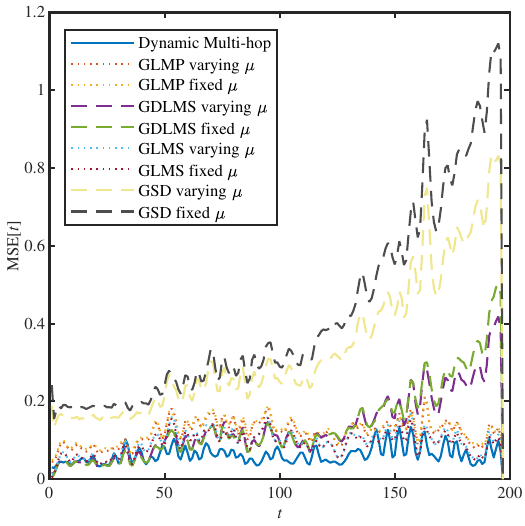}         \vspace{-20 pt}
		\caption*{Normal  Subject, SNR = $10$.}
	\end{subfigure}
	\begin{subfigure}{0.32\textwidth}
		\centering
		\includegraphics[trim= 180 270 180 270,clip, width=\textwidth]{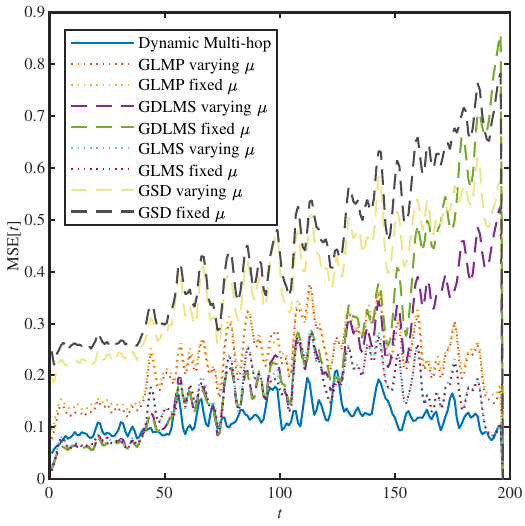}
		\vspace{-20 pt}
		\caption*{Autistic  Subject, SNR = $3$.}
	\end{subfigure}
	\begin{subfigure}{0.32\textwidth}
		\centering
		\includegraphics[trim= 180 270 180 270,clip, width=\textwidth]{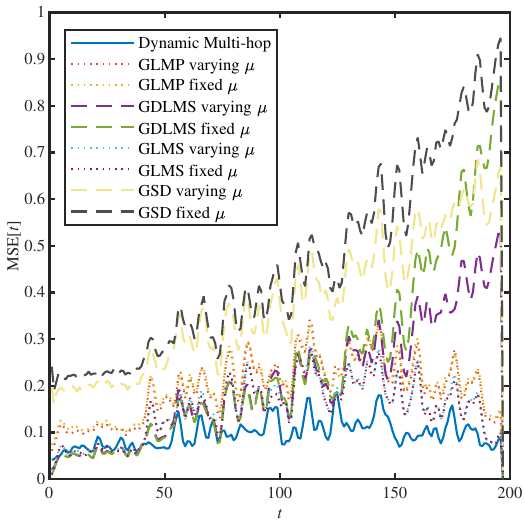}
		\vspace{-20 pt}
		\caption*{Autistic  Subject, SNR = $5$.}
	\end{subfigure}
	\begin{subfigure}{0.32\textwidth}
		\centering
		\includegraphics[trim= 180 270 180 270,clip, width=\textwidth]{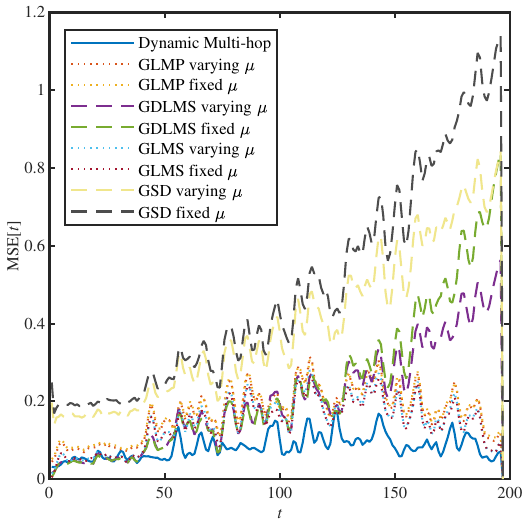}         \vspace{-20 pt}
		\caption*{Autistic  Subject, SNR = $10$.}
	\end{subfigure}
	\caption{The MSE$[t]$ of the online estimations of brain ROI signals for normal and autistic subjects under various SNR settings for Dynamic Multi-hop estimation and adaptive GSP algorithms. Dashed lines: spatial methods, dotted lines: spectral methods.}
	\label{fig_brain_MSE}
\end{figure*}

\section{Applying Dynamic Multi-hop to Brain Networks}
\label{sec_brain}
In this section, we explore the application of the Dynamic Multi-hop framework within the context of functional brain connectivity analysis, demonstrating its capability to model and analyze real-world time-varying interactions.
In short, functional brain connectivity analysis interprets connections through brain regions in terms of the statistical dependencies between signals emitting from distinct regions of the brain. 
Employing graph-based methods to assess brain connectivity opens up innovative avenues for examining how functional deficits in the brain correlate with structural disruptions implicated in various brain disorders \cite{zhao_2024_sequential}. 
Nevertheless, it is important to acknowledge that assumptions of noiseless data and unchanging graph topologies hinder the effectiveness of current methodologies.

\subsection{Dataset and Preprocessing}
\label{sec_experiment_brain_pre}
Dynamic Multi-hop will be applied to the normal and Autistic subjects selected from the Autism Brain Imaging Data Exchange (ABIDE) dataset, in which the fMRI data was gathered from different internationally shared open databases \cite{craddock2013neuro}. 
The time-varying node signals $\boldsymbol{x}_g[t]$ are provided directly from the data in the form of a multivariate time series, which was formulated by brain regions that are identified from fMRI images. 
Connectivity within the brain can be measured by transforming neighborhood relationships into a connectivity matrix, where each row and column represents different Regions of Interest (ROI) in the brain \cite{li2021braingnn}.
In the experiments, we selected the ROI division which follows the Harvard-Oxford atlas, giving us $111$ different regions corresponding to $N_n = 111$ nodes. 
The signals within each ROI will naturally form a time-varying node signal $\boldsymbol{x}_g[t]$. 
We selected a normal subject and an Autistic subject within the ABIDE dataset, all with $T = 196$ time steps. 
The noise on the node signals is modeled by Gaussian distribution with zero mean with the variance set corresponding to Signal-to-Noise Ratio (SNR) of 3, 5, and 10. 

We will conduct an online estimation of the ground truth signal $\boldsymbol{x}_g[t]$ under the setting where there will be additive noise and randomly missing node observations.
In this experiment, the selected data is further split into training ($t = [1, 40]$), validation ($t = [41, 60]$), and testing ($t = [61, T]$) sets.
Additionally, we will use Dynamic Multi-hop to reveal a series of dynamic graph topologies from the multivariate ROI time series $\boldsymbol{x}_g[t]$ to represent the dynamic interactions among different brain regions. 
Due to the significant range differences among the signals in each node, we normalized each of the nodes using their means in the training set.

The initial static graph $\mathcal{G}$ is formed by applying SGM in which the correlation matrix of the training multivariate time series of ROI is pruned.
We employ a combination of the threshold-based method, similar to the approach in \cite{li2021braingnn}, and the Planar Maximally Filtered Graph \cite{tumminello2005tool}.
The correlation among $111$ nodes is calculated across $196$ time steps and recorded in the correlation matrix $\mathbf{C}_F$.
Using all the $N_n\times N_n$ elements in $\mathbf{C}_F$, we define a graph adjacency matrix $\mathbf{A_F}$, where the $ij^{th}$ non-diagonal elements are $\mathbf{A}_{F,ij} = $ abs$($sign$(\mathbf{C}_{F,ij}))$ and the $i^{th}$ diagonal elements are $\mathbf{A}_{F,ii} = 0$.
To formulate sparse edge connections, we connect each node only with its top-$3$ nodes showing the largest absolute correlations. 
Additionally, we link nodes that have an absolute correlation exceeding a threshold of $0.95$.
The choice of using the top-$3$ most (absolute) correlated nodes is done by following the suggestion in \cite{tumminello2005tool}. 
Viewing the absolute correlation from another aspect, we maintained both data homophily (positive correlation) and data heterophily (negative correlation) \cite{zhu2021graph_heterophily}. 
The formulated sparse graph with fixed topology will be denoted as $\mathcal{G}$ with edge set is $\mathcal{E} = \{ e_1 \dots e_{N_e} \}$. 
A demonstration of the graph $\mathcal{G}$ constructed on a normal subject and an Autistic subject is shown in Figure~\ref{fig_brain_total_graph} using the same topology-forming techniques and the same parameters.

\begin{figure}[htb]
	\centering
	\includegraphics[width=0.45\textwidth]{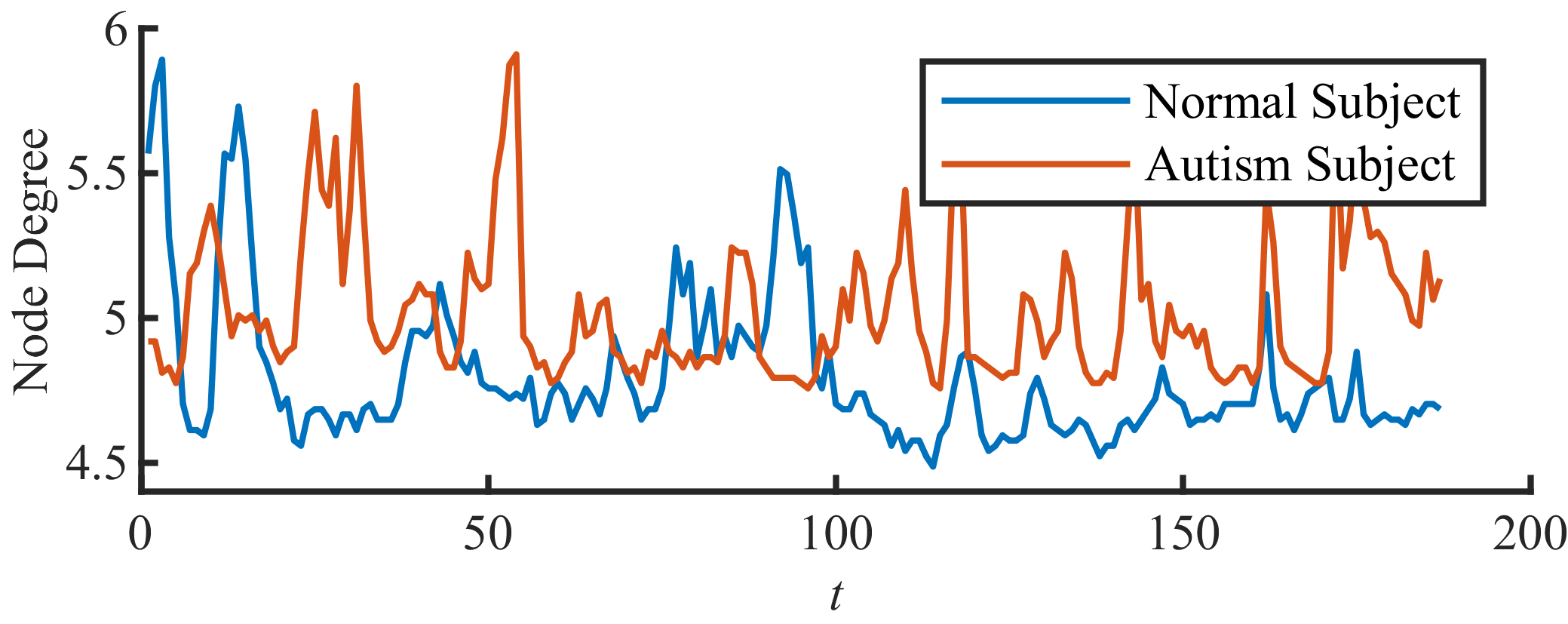}
	\caption{The average unweighted node degree over time.}
	\label{fig_node_degree}
\end{figure}

\begin{figure}[htb]
	\centering
	\begin{subfigure}{0.45\textwidth}
		\includegraphics[trim = 0 10 0 0,width = \textwidth]{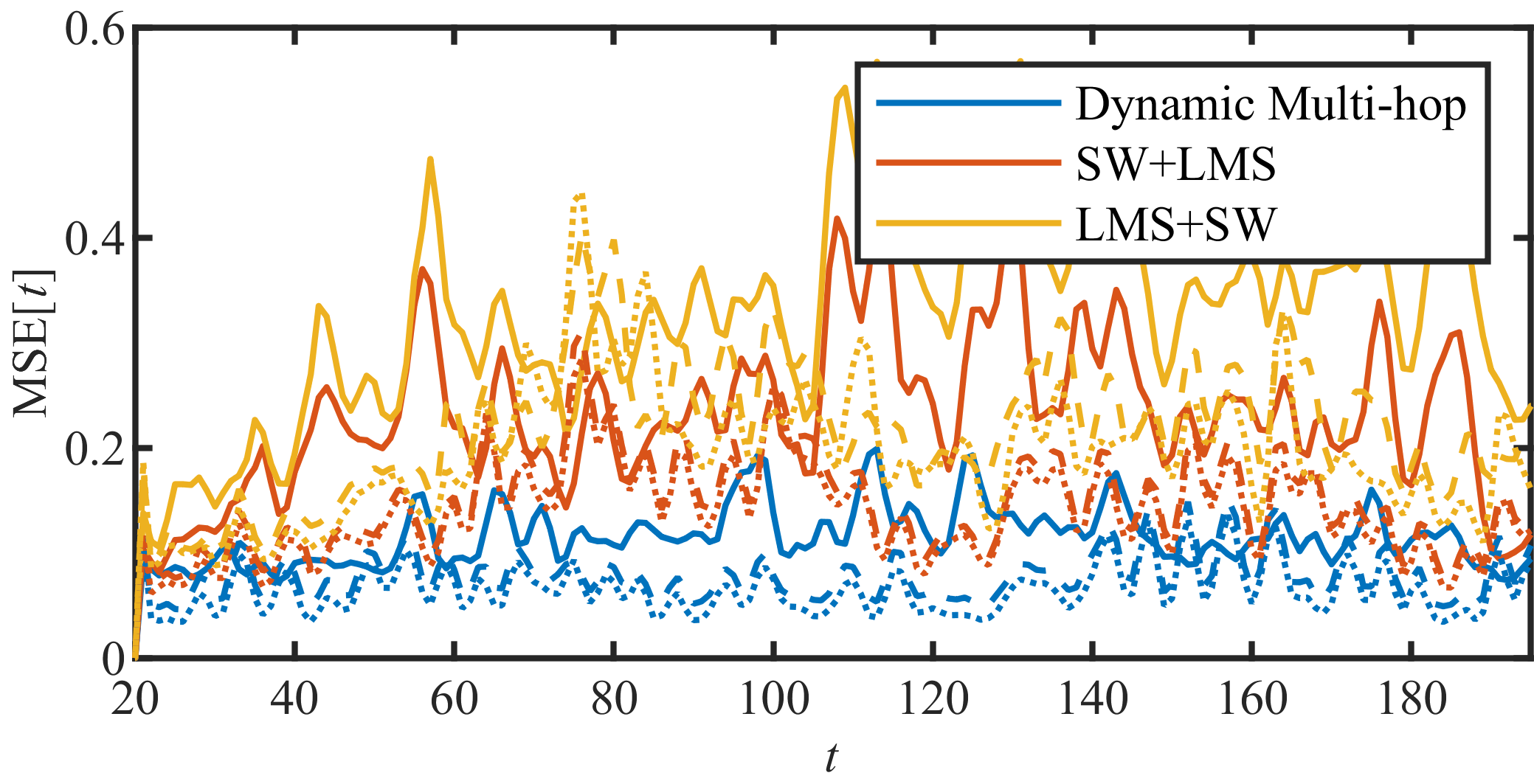}
		\caption*{Normal subject}
	\end{subfigure}
	\begin{subfigure}{0.45\textwidth}
		\includegraphics[trim = 0 10 0 0,width = \textwidth]{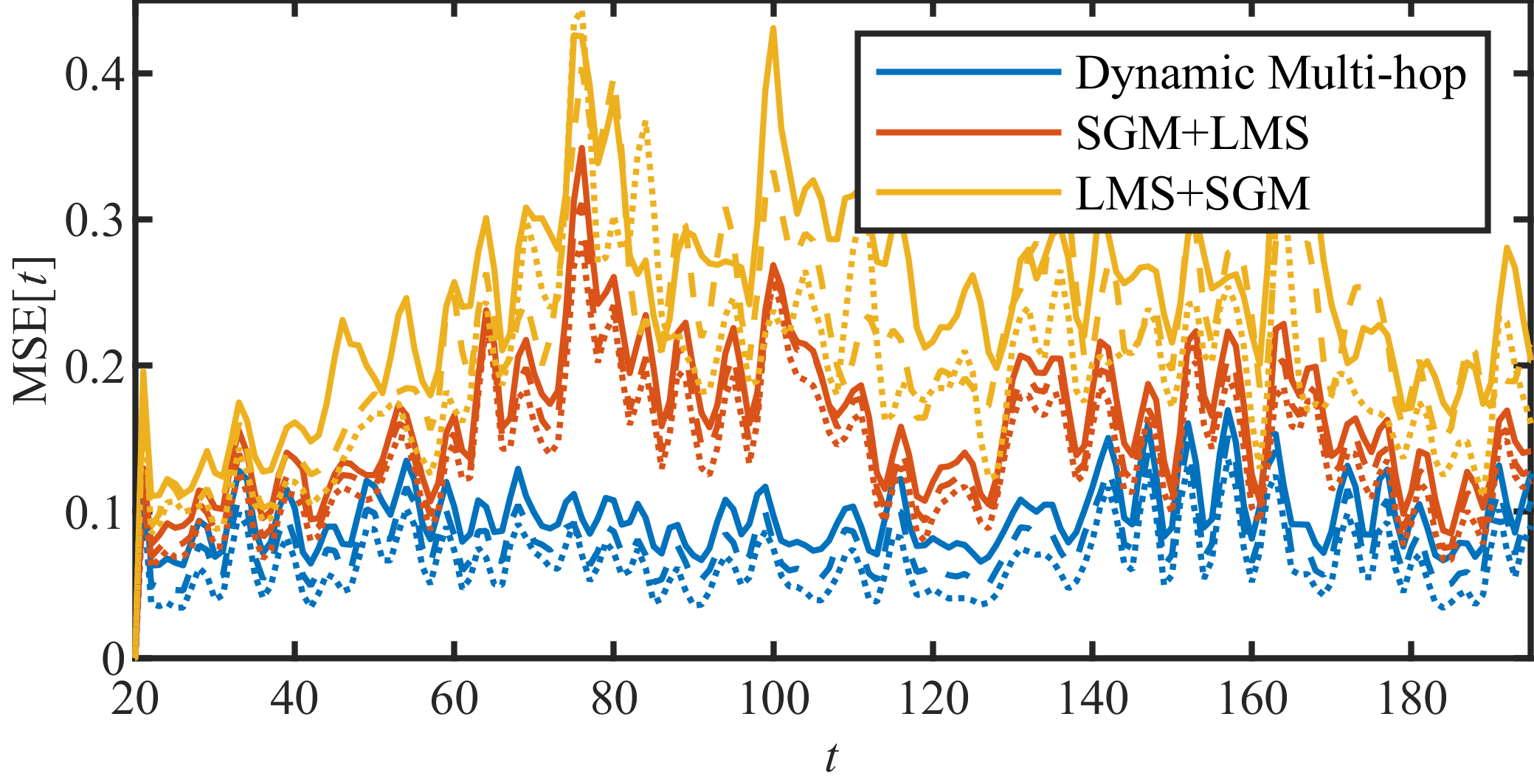}
		\caption*{Autistic subject}
	\end{subfigure}
	\caption{The MSE$[t]$ of the online estimations under various SNR settings for Dynamic Multi-hop estimation compared against the combination of GLMS and SGM. Solid lines: SNR $= 3$, dashed lines: SNR$ = 5$, dotted lines: SNR$ = 10$.}
	\label{fig_slide}
\end{figure}

\subsection{Dynamic Multi-hop Estimation on Brain Networks}
We would like to use Dynamic Multi-hop to form dynamic brain networks from the time-varying node signals and the initial static graph while we will conduct online estimations of the node signal from partial observations containing noise using Dynamic Multi-hop estimation in Algorithm~\ref{alg_dynamic_multi_hop_adaptive}. 
To put it another way, our Dynamic Multi-hop estimation will be reconstructing a time-varying signal by completing the tasks of online estimation, denoising, and transductive learning at the same time by enhancing a static graph into a series of dynamic graphs.
The estimation process in Algorithm~\ref{alg_dynamic_multi_hop_adaptive} is iterative and alternates between forming a dynamic topology and outputting online estimation, so the Dynamic Multi-hop estimation should be capable of effectively completing both tasks.

The Dynamic Multi-hop estimation in Algorithm~\ref{alg_dynamic_multi_hop_adaptive} will be compared against the following adaptive GSP algorithms at the online estimation of time-varying node signals: GLMS \cite{bib_LMS}, GDLMS \cite{Roula_2017_LMS_Diffusion}, GLMP \cite{nguyen2020_LMP}, G-Sign \cite{yan_2022_sign}, and GSD \cite{yan_2023_sign}. 
For the hyperparameter setting, the number of hops $P$ is set to $P=6$.
For the step size $\mu$ in \eqref{eq_update}, the Dynamic Multi-hop will adopt a time-varying step size that changes with respect to the estimation residual $\mathbf{M}{\boldsymbol{y}[t]-\hat{\boldsymbol{x}}[t]}$:
\begin{equation}
	\mu = (\mu_{\text{max}} - \mu_{\text{min}}) \cdot e^{-\|\mathbf{M}(\boldsymbol{y}[t]-\hat{\boldsymbol{x}}[t])\|_2} + \mu_{\text{min}},
	\label{eq_mu}
\end{equation}
where $\mu_{\text{max}}$ is the maximum possible value of $\mu$ and $\mu_{\text{min}}$ is the minimum possible value of $\mu$.
The baseline algorithms are tested for both cases for adopting fixed step sizes and time-varying step sizes.
The step size is set to $\mu_{\text{min}} = 0.8$ and $\mu_{\text{min}} = 3.5$.
For the baselines that use fixed step size, $\mu = 0.9$ is chosen because it is the average $\mu$ that appeared when operating on the training set.
All the tested algorithms will use an ideal low pass filter with a pass band of [0, $0.4\lambda_{N_n}$].

To generate the time-varying signals on the edges within the training set, we will use a sliding window of length ${\omega_{sl}} = 10$ with stride size $1$ to calculate a series of sliding window correlations $\mathbf{C}[t]$ from the past signals $\boldsymbol{x}_g[t]$ to $\boldsymbol{x}_g[t-\omega_{sl}]$. 
However, initially, we only keep the entries corresponding to the edges within the edge set $\mathcal{E}$ to enforce sparsity. 
Again, we take the absolute value of the correlation matrices $\mathbf{C}[t]$ to avoid negative edge weights, resulting in time-varying edge weights (edge signal).
For every edge in the static topology of $\mathcal{G}$, the corresponding edge signals are mapped onto $\mathcal{G}$ by replacing the corresponding elements in the adjacency matrix $\mathbf{A}$ with the signal elements in $\boldsymbol{w}$. 
It should be emphasized that in this case, the signals on the edges are fixed but the signals on the nodes are time-varying. 

In the testing phase, the edge signals will also be estimated using the correlation method but will be done based on the node signal estimations $\hat{\boldsymbol{x}}[t]$ instead of the ground truth $\boldsymbol{x}_g[t]$ because only partial observations of the node signal are available.
An illustration of the formed dynamic graphs on the testing set is shown in Figure~\ref{fig_brain_prune}.
The online estimation results of the node signals that reflect the brain ROI activities will be evaluated by the MSE$[t]$ defined as
\begin{equation}
	\text{MSE}[t] = \frac{1}{{N_n}R}\sum_{r=1}^{R}\sum_{i=1}^{N_n}{({x}_{g,i,r}[t]-\hat{{x}}_{i,r}[t])^2}.
	\label{eq_MSE}
\end{equation}
The MSE$[t]$ results are shown in Figure~\ref{fig_brain_MSE} for SNR = $3, 5, $ and $10$. 
The experiment will be repeated for $R = 100$ runs.

Comparing the graphs at time $t = 52$ with $t = 60$, one can observe that there is not only a change in the signals on the edges but also there are dynamic edge connections formed by our Dynamic Multi-hop approach. 
This is more apparent when comparing Figure~\ref{fig_brain_total_graph} with Figure~\ref{fig_brain_prune}, in which different edges are formed on top of the given static topology when $t = 52$ with $t = 60$.
We also give a visual comparison to show the difference between the graph topology captured by Dynamic Multi-hop and the original graph in Figure~\ref{fig_change}.
To further confirm that Dynamic Multi-hop provides us with a dynamic topology over time, we will calculate the node degree.
So, at each time instance, the unweighted node degree will give us the number of edges connected to that node. 
If this statistic changes at each time instance, then we can confirm that the number of edges that each node has is also changing.
We plotted the average (unweighted) node degree over time in Figure~\ref{fig_node_degree}. 
Visually analyzing Figure~\ref{fig_node_degree}, we can confirm that for both types of subjects, the average unweighted node degree indeed changes over time, which confirms the fact that Dynamic Multi-hop generates dynamic graph topologies.

Looking at the MSE$[t]$ performance of all the tested algorithms in Figure~\ref{fig_brain_MSE}, we can observe that the Dynamic Multi-hop Estimation has the lowest MSE$[t]$ among all the tested algorithms under all the noise scales at most time instances. 
The reason that Dynamic Multi-hop is able to have relatively low MSE$[t]$ is due to the fact that it formulates and utilizes dynamic topologies.
The baseline algorithms that use the static topology obtained from the training set, which are from earlier time instances, no longer perform well in the testing set because the topology obtained in the training set no longer reflects the dynamics of the nodes in later time instances.
On the other hand, the Dynamic Multi-hop forms accurate estimations because it is able to reveal the latent edges from the node signals.
The low MSE$[t]$ confirms that these revealed edges accurately represent the interactions among the node signals previously absent in the initial static topology defined by the training set.
Additionally, the Dynamic Multi-hop estimation shown in Algorithm~\ref{alg_dynamic_multi_hop_adaptive} effectively uses these newly formed edges in the resulting dynamic graphs to define the filtering operation shown in \eqref{eq_adaptive_filter} to conduct online estimation. 

The best-performing baseline algorithm in Figure~\ref{fig_brain_MSE}, GLMS with fixed step size, is combined with dynamic graphs formed by applying sliding window SGM method shown in \cite{zhao_2024_sequential} to compare against the Dynamic Multi-hop estimation. 
Doing so will allow the Dynamic Multi-hop to be compared against algorithms that use dynamic graphs instead of static graphs. 
The estimation of GLMS will undergo two different settings: 1. use GLMS to estimate the node signal first, then use SGM to form the dynamic topology (denoted as GLMS+SGM in Figure~\ref{fig_slide}); 2. use SGM to form the dynamic topology first, then use GLMS to estimate the node signals (denoted as SGM+GLMS in Figure~\ref{fig_slide}). 
The MSE$[t]$ results of noise settings with SNR = $3, 5, $ and $10$ are shown in Figure~\ref{fig_slide} for both normal and autistic subjects. 
Visually inspecting Figure~\ref{fig_slide}, one can notice that the Dynamic Multi-hop estimation again has lower MSE$[t]$ even when comparing with the combination of SGM and GLMS.
From the methodology perspective, both the Dynamic Multi-hop and the SGM GLMS combination use SGM to formulate dynamic graphs by thresholding the edge weights and then pruning the edges.
However, the Dynamic Multi-hop is able to utilize the power of GSP and TSP, which allows it to keep the edges that represent not only the 1-hop node interactions but also strong node interactions from further node neighbors. 
Additionally, the Dynamic Multi-hop is designed to consider the scenario of noisy and partial observations in the node signals.
The limitation of the combination of SGM and GLMS is that the SGM assumes the knowledge of clean full observations on the nodes to form graph topologies, but in reality, clean full observation of the data is sometimes not obtainable. 
So, compared to Dynamic Multi-hop, the SGM and GLMS combination is not as suitable as Dynamic Multi-hop at conducting online estimation under noise and partial observation. 

\begin{figure}[htb]
	\centering
	\begin{subfigure}{0.26\textwidth}
		\includegraphics[trim= 0 0 35 0,clip, width = \textwidth]{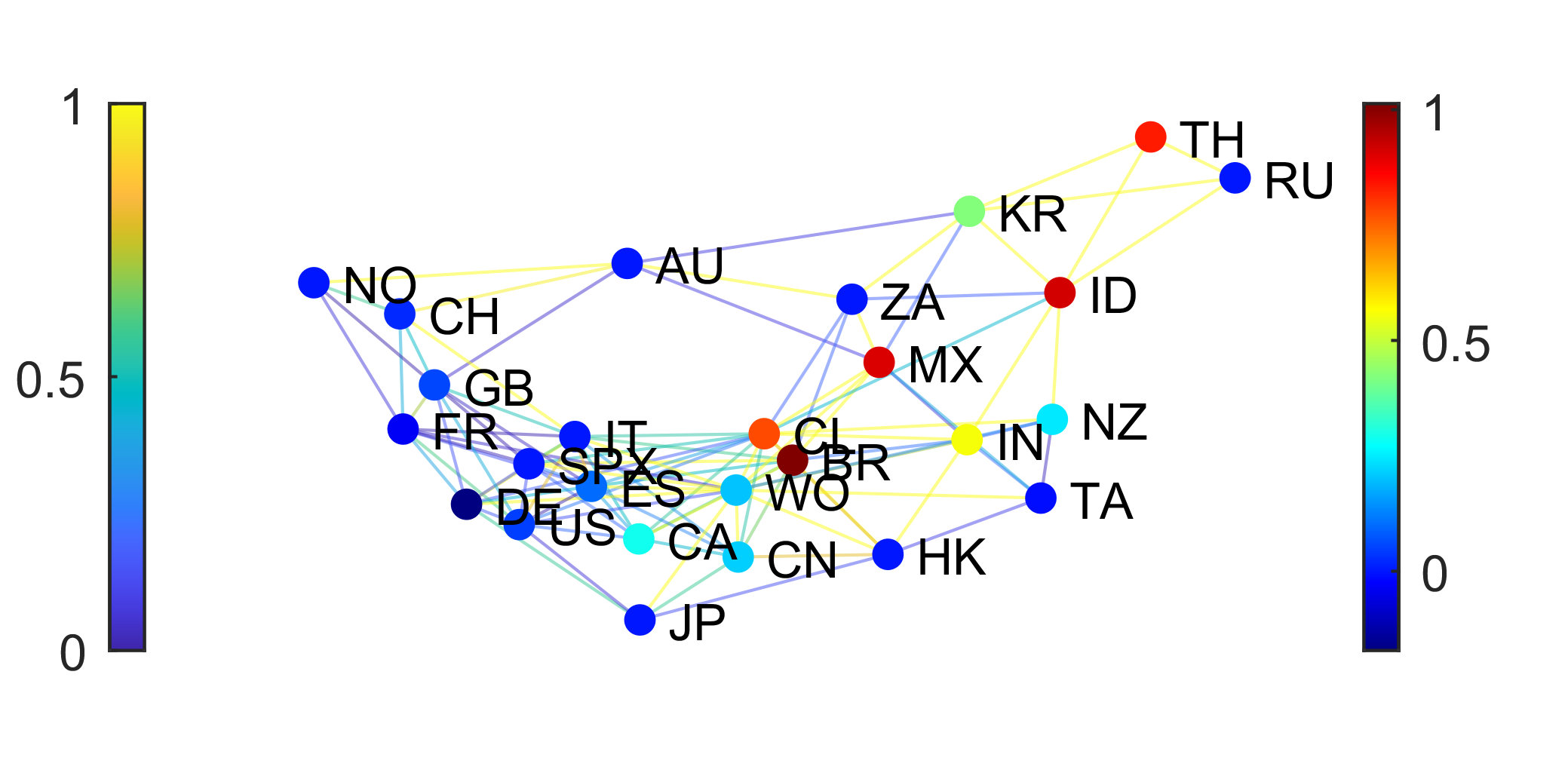}
		\vspace{-30 pt}
		\caption*{$t = 300$}
	\end{subfigure}
	\begin{subfigure}{0.22\textwidth}
		\includegraphics[trim= 35 0 35 0,clip, width = \textwidth]{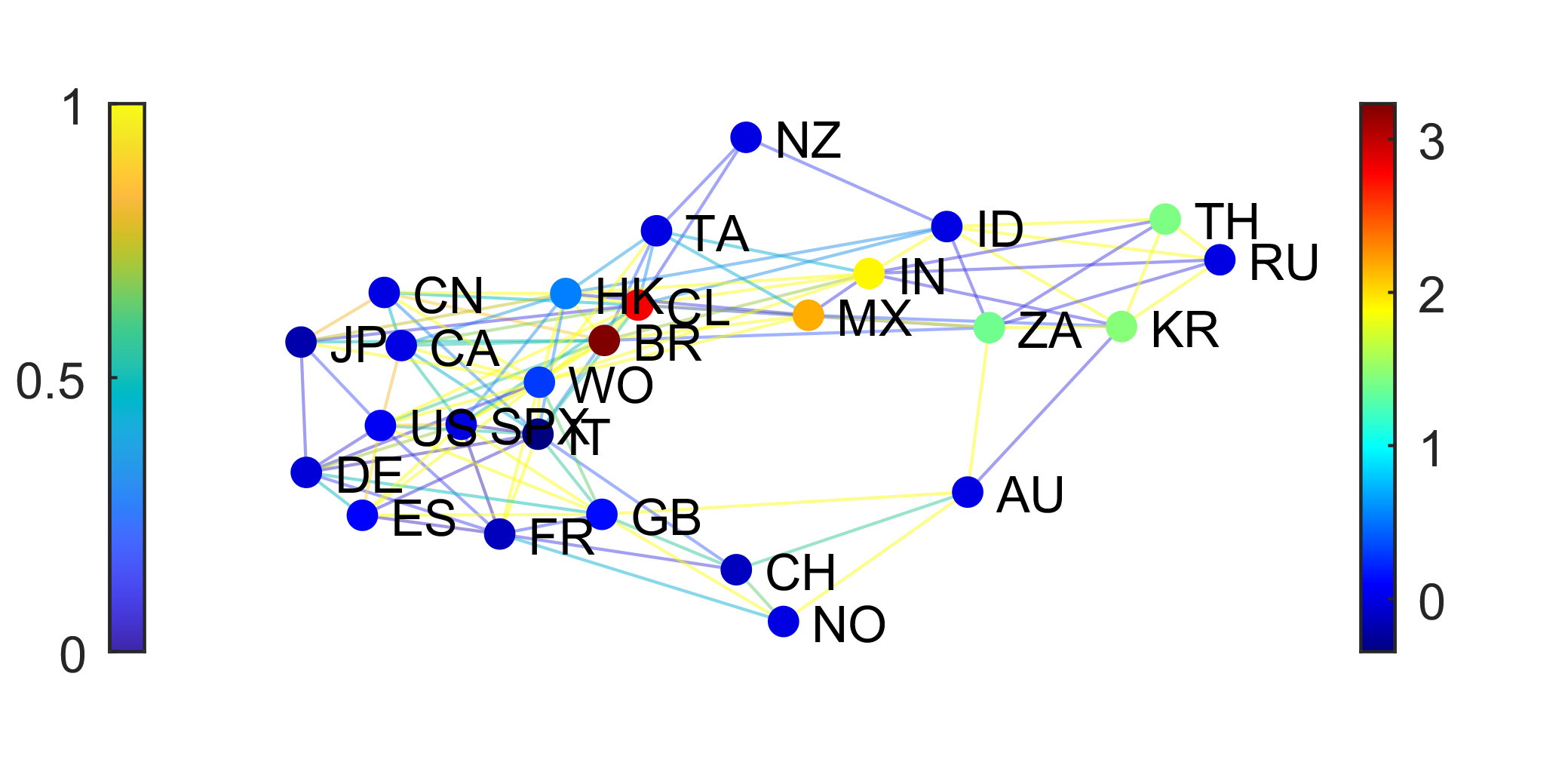}
		\vspace{-30 pt}
		\caption*{$t = 600$}
	\end{subfigure}
	\caption{Illustration of the graph topologies formed by Dynamic Multi-hop on the MSCI dataset at two different time instances.}
	\label{fig_stock_tops}
\end{figure}

\begin{figure}[htb]
	\centering
	\includegraphics[trim = 190 305 190 305, width=0.4\textwidth]{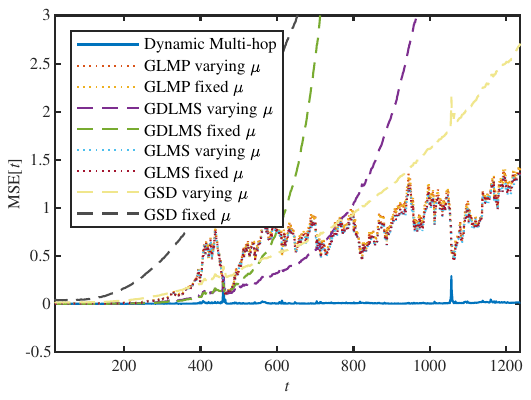}
	\caption{The MSE$[t]$ of the online estimations of stock closing MSCI indices. Dashed lines: spatial methods, dotted lines: spectral methods.}
	\label{fig_MSE_stock_static}
\end{figure}

\begin{figure}[htb]
	\centering
	\includegraphics[width=0.45\textwidth]{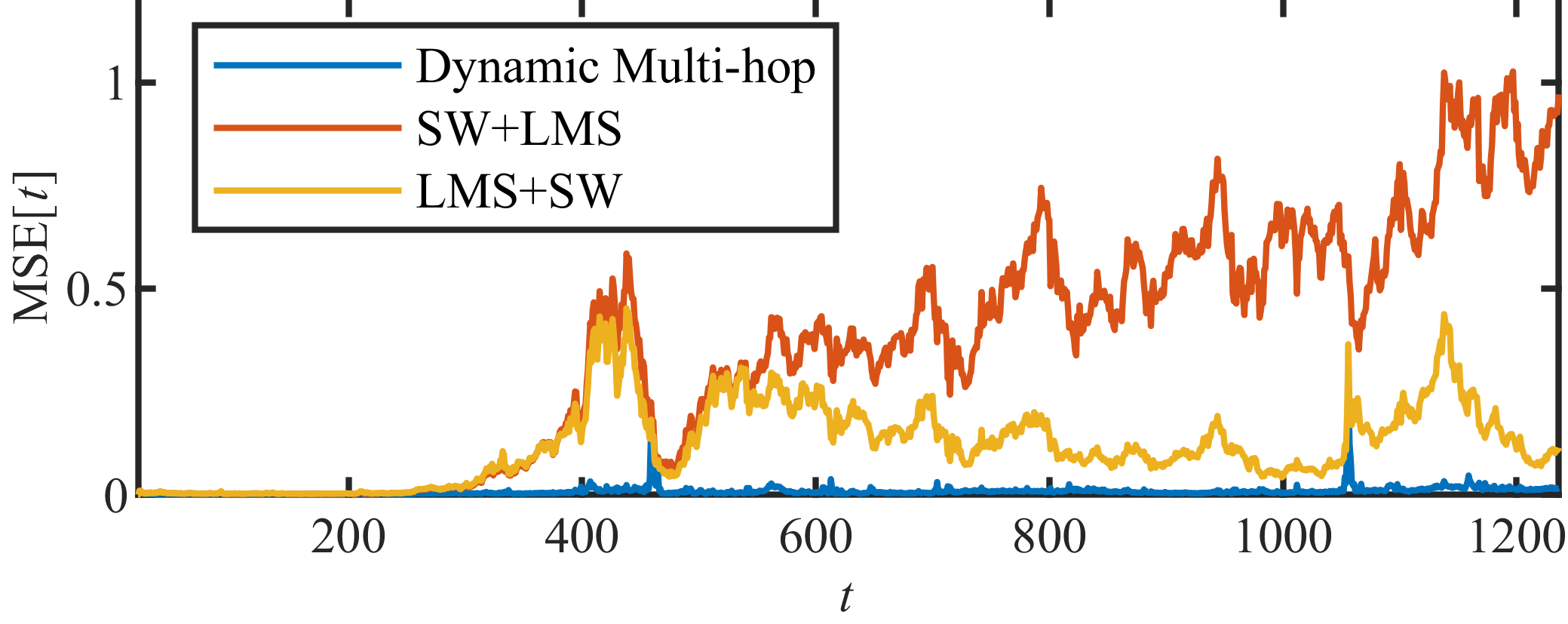}
	\caption{The MSE$[t]$ of the online estimations of stock closing MSCI indices of algorithms using dynamic graph topologies.}
	\label{fig_MSE_stock_slide}
\end{figure}

\begin{figure}[h]
	\centering
	\includegraphics[width=0.45\textwidth]{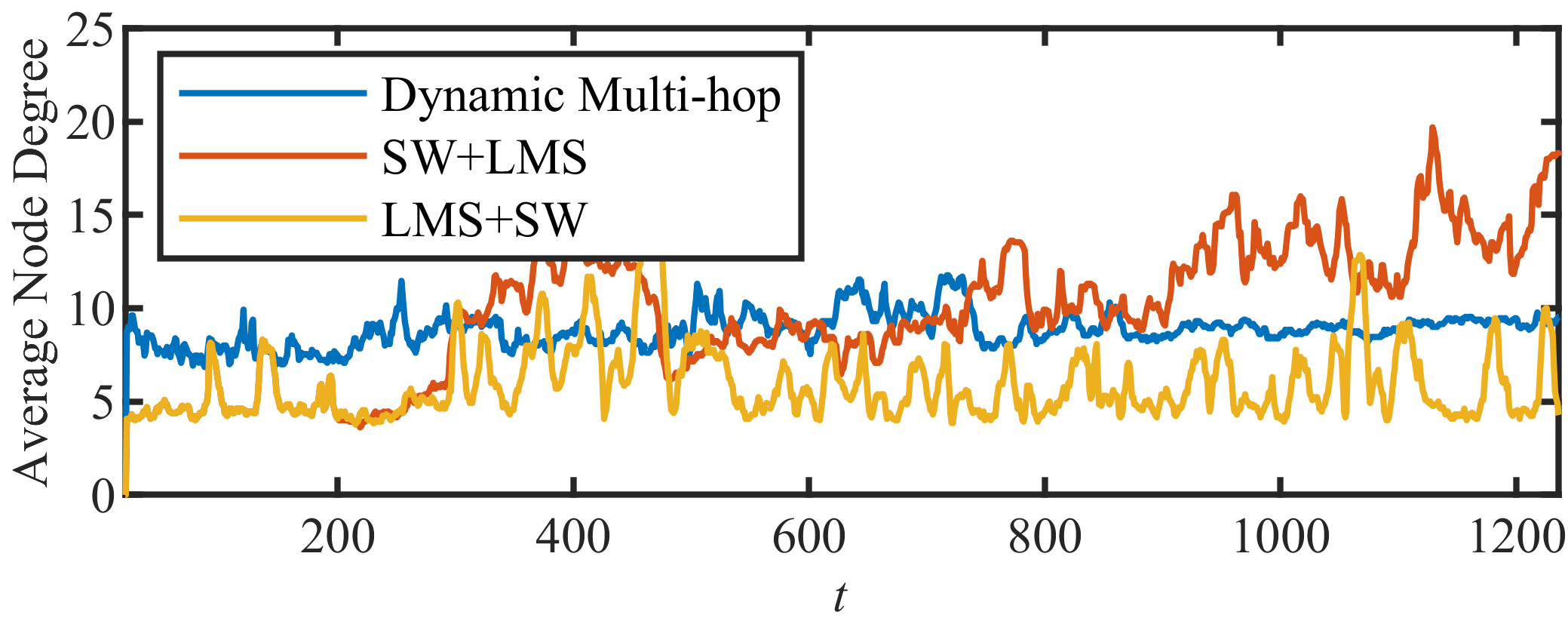}
	\caption{The average unweighted node degree of algorithms using dynamic graph topologies.}
	\label{fig_stock_deg}
\end{figure}

\section{Dynamic Multi-hop for Stock Market Analysis}
\label{sec_stock}
In this experiment, the Dynamic Multi-hop Estimation in Algorithm~\ref{alg_dynamic_multi_hop_adaptive} is applied to a financial dataset consisting of the Morgan Stanley Capital International (MSCI) indices closing prices from 26 global stock markets, covering the period from 2020 to 2023, with a total of $T=1238$ time points \cite{2024_Stable_Muvunza}.
In this real-world dataset, each stock is represented by 1 node in a graph and the MSCI indices closing price is a multi-variate time series and will be used as the node signals $\boldsymbol{x}[t]$. 
We divided the dataset so the training set comprises the interval $t = [1, 200]$, the validation set is the interval $t = [201, 400]$, and the testing set is the interval from $t = [401, T]$.
The initial static graph and the time-varying edge weights $\boldsymbol{w}[t]$ are obtained using the exact same approach discussed in Section~\ref{sec_experiment_brain_pre}. 
The dataset also went through the same preprocessing procedure as seen in Section~\ref{sec_experiment_brain_pre}.
The node signal is corrupted by the additive Gaussian noise with SNR $ = 3$, with $30$ percent missing observations.
Our Dynamic Multi-hop estimation aims to reconstruct the time-varying MSCI indices using Algorithm~\ref{alg_dynamic_multi_hop_adaptive}.

Dynamic Multi-hop Estimation in Algorithm~\ref{alg_dynamic_multi_hop_adaptive} is compared to GLMS, GDLMS, GLMP, G-Sign, GSD, GLMS+SGM and SGM+GLMS. 
The fixed step size is set to be $\mu = 0.4$, and the time-varying step size is determined using the method shown in \eqref{eq_mu} with $\mu_{\text{min}} = 0.2$ and $\mu_{\text{min}} = 0.6$. 
Two illustrations of the graphs generated using Dynamic Multi-hop are shown in Figure~\ref{fig_stock_tops}. 
The MSE$[t]$ results of Dynamic Multi-hop compared against adaptive GSP methods and dynamic methods are shown in Figure~\ref{fig_MSE_stock_static} and Figure~\ref{fig_MSE_stock_slide} respectively.
The unweighted average node degree of the dynamic graphs in the dynamic methods is shown in Figure~\ref{fig_stock_deg}.

First of all, it can be confirmed that the Dynamic Multi-hop Estimation is able to enhance the static graph into dynamic graphs by both observing the graph topologies as shown in Figure~\ref{fig_stock_tops} and analyzing the time-varying node degree at each time instance in Figure~\ref{fig_stock_deg}.
Now, looking at the MSE$[t]$, it is clear that the Dynamic Multi-hop is able to form the lowest MSE $[t]$ among all the tested algorithms. 
For the problem setting considered in this experiment, for an online estimation to be accurate and to avoid error cumulation, it is required that the algorithms deployed should efficiently utilize the graph topology to accurately reconstruct the actual signal from the missing and noisy observation.
Analyzing Figure~\ref{fig_MSE_stock_static}, the MSE$[t]$ diverges for the algorithms that only use the initial static topology obtained from the training set because the static topology is unable to capture the dynamic interactions among the time-varying node signals.
This divergent behavior is observed regardless of whether the algorithm approaches this task from a spectral point of view or a spatial point of view. 
On the other hand, Dynamic Multi-hop estimation produces accurate reconstruction results, indicating that the dynamic topologies accurately capture the dynamics of the data on the nodes and is able to utilize this dynamic. 
Inspecting Figure~\ref{fig_MSE_stock_slide} and Figure~\ref{fig_stock_deg} together, we see that the Dynamic Multi-hop outputs dynamic graphs with a relatively steady average node degree for each node, and the average node degree lies in-between SW+LMS ans LMS+SW while maintaining a lower MSE$[t]$. 
This indicates that the graph topologies of Dynamic Multi-hop have high representation power which more accurately captures the time-varying interactions among different stocks while maintaining relatively good sparsity.

\section{Conclusion and Future Work}

The Dynamic Multi-hop is proposed for constructing dynamic graphs from a static graph to represent the interactions among the time-varying signals on the graph nodes.
Leveraging the obtained dynamic graphs, we are able to conduct effective online estimations of time-varying signals on the graph nodes under noisy and missing observations compared to previous online graph algorithms. 
We confirmed that the Dynamic Multi-hop model not only outputs dynamic graphs but also obtains more accurate online estimations when applying the Dynamic Multi-hop on brain and stock networks. 

Additional improvements can be explored to enhance and extend the Dynamic Multi-hop Model. 
Further improvements can be made to enhance and extend the Dynamic Multi-hop Model with advanced methodologies such as time-series analysis techniques and state-space models, which have demonstrated impressive performance both independently and in graph-based formulations \cite{SAFFARIMIANDOAB2021115197, Isufi2019gvarma, francq2019garch, kalman_1960_new, Sagi_2023_graph_EFK_Kalman, alippi2023graph_kalman_new, costagli2007image}."
Non-linearity could be introduced into the Dynamic Multi-hop Model by replacing the graph adaptive filters with Graph Neural Network (GNN) components. GNNs are well-suited for capturing complex relationships in graph-structured data and have demonstrated exceptional performance in various machine learning tasks such as classification, regression, and clustering \cite{kipf2016semi}.
These extensions will allow Dynamic Multi-hop to make predictions under more complicated scenarios and predict further time steps.
Currently, we model the noise using Gaussian distributions. In future work, we aim to evaluate the performance of the Dynamic Multi-hop Model under $\alpha$-stable distributions. 
As a generalization of Gaussian distributions, $\alpha$-stable distributions are well-suited for capturing impulsive, heavy-tailed behaviors \cite{kuruoglu1997new, herranz2004alpha}. 
The robustness of the Dynamic Multi-hop can potentially be improved by the consideration of impulsiveness through $\alpha$-stable noise models.
Furthermore, the edge pruning based on the SGM can be further improved by more advanced methodologies of edge pruning to produce topologies that can capture the dynamics in the nodes with even higher representation power.
Lastly, would also like to extend the fields of applications of our Dynamic Multi-hop to solve additional real-world problems, such as in material science \cite{SAFFARIMIANDOAB2021115197} and image separation \cite{costagli2007image}.

\bibliographystyle{IEEEbib}
\bibliography{paper}

\begin{thebibliography}{10}

\bibitem{Ortega_graph_2018}
A.~Ortega, P.~Frossard, J.~Kovačević, J.~M.~F. Moura, and P.~Vandergheynst,
\newblock ``Graph signal processing: Overview, challenges, and applications,''
\newblock {\em Proceedings of the IEEE}, vol. 106, no. 5, pp. 808--828, 2018.

\bibitem{Dong_Graph_ML_2020}
X.~Dong, D.~Thanou, L.~Toni, M.~Bronstein, and P.~Frossard,
\newblock ``Graph signal processing for machine learning: A review and new perspectives,''
\newblock {\em IEEE Signal Processing Magazine}, vol. 37, no. 6, pp. 117--127, 2020.

\bibitem{Leus_2024_GSP_review}
G.~Leus, A.~G. Marques, J.~M. Moura, A.~Ortega, and D.~I. Shuman,
\newblock ``Graph signal processing: History, development, impact, and outlook,''
\newblock {\em IEEE Signal Processing Magazine}, vol. 40, no. 4, pp. 49--60, 2023.

\bibitem{bib_LMS}
P.~D.~Lorenzo, S.~Barbarossa, P.~Banelli, and S.~Sardellitti,
\newblock ``Adaptive least mean squares estimation of graph signals,''
\newblock {\em IEEE Transactions on Signal and Information Processing over Networks.}, vol. 2, no. 4, pp. 555 -- 568, 2016.

\bibitem{Roula_2017_LMS_Diffusion}
R.~Nassif, C.~Richard, J.~Chen, and A.~H. Sayed,
\newblock ``A graph diffusion lms strategy for adaptive graph signal processing,''
\newblock in {\em Asilomar}, 2017, pp. 1973--1976.

\bibitem{yan_2023_sign}
Y.~Yan and E.~E. Kuruoglu,
\newblock ``Fast and robust wind speed prediction under impulsive noise via adaptive graph-sign diffusion,''
\newblock in {\em IEEE CAI}, 2023, pp. 302--305.

\bibitem{lu2024latent}
J.~Lu, Y.~Xu, H.~Wang, Y.~Bai, and Y.~Fu,
\newblock ``Latent graph inference with limited supervision,''
\newblock {\em Advances in Neural Information Processing Systems}, vol. 36, 2024.

\bibitem{de2022latent}
H.~S. de~Oc{\'a}riz~Borde, A.~Kazi, F.~Barbero, and P.~Lio,
\newblock ``Latent graph inference using product manifolds,''
\newblock in {\em ICLR}, 2022.

\bibitem{kazi2022differentiable}
A.~Kazi, L.~Cosmo, S.-A. Ahmadi, N.~Navab, and M.~M. Bronstein,
\newblock ``Differentiable graph module for graph convolutional networks,''
\newblock {\em IEEE Transactions on Pattern Analysis and Machine Intelligence}, vol. 45, no. 2, pp. 1606--1617, 2022.

\bibitem{Panagopoulos2021pandemic}
G.~Panagopoulos, G.~Nikolentzos, and M.~Vazirgiannis,
\newblock ``Transfer graph neural networks for pandemic forecasting,''
\newblock {\em AAAI}, vol. 35, no. 6, pp. 4838--4845, May 2021.

\bibitem{Kuruoglu_2016_gene}
S.~Ancherbak, E.~E. Kuruoglu, and M.~Vingron,
\newblock ``Time-dependent gene network modelling by sequential monte carlo,''
\newblock {\em IEEE/ACM Transactions on Computational Biology and Bioinformatics}, vol. 13, no. 6, pp. 1183--1193, 2016.

\bibitem{2024_Market_Qin}
D.~Qin and E.~E. Kuruoglu,
\newblock ``Graph learning based financial market crash identification and prediction,''
\newblock in {\em IEEE CAI}, 2024.

\bibitem{lauritzen1996graphical}
S.~L. Lauritzen,
\newblock {\em Graphical models}, vol.~17,
\newblock Clarendon Press, 1996.

\bibitem{Altenbuchinger_2020_GMM}
M.~Altenbuchinger, A.~Weihs, J.~Quackenbush, H.~J. Grabe, and H.~U. Zacharias,
\newblock ``Gaussian and mixed graphical models as (multi-)omics data analysis tools,''
\newblock {\em Biochimica et Biophysica Acta - Gene Regulatory Mechanisms}, vol. 1863, no. 6, pp. 194418, 2020.

\bibitem{yuan2007_GMM_model}
M.~Yuan and Y.~Lin,
\newblock ``Model selection and estimation in the gaussian graphical model,''
\newblock {\em Biometrika}, vol. 94, no. 1, pp. 19--35, 2007.

\bibitem{dong_2016_learning_laplacian}
X.~Dong, D.~Thanou, P.~Frossard, and P.~Vandergheynst,
\newblock ``Learning laplacian matrix in smooth graph signal representations,''
\newblock {\em IEEE Transactions on Signal Processing}, vol. 64, no. 23, pp. 6160--6173, 2016.

\bibitem{sardellitti_2021_online_small_pertubation}
S.~Sardellitti, S.~Barbarossa, and P.~Di~Lorenzo,
\newblock ``Online learning of time-varying signals and graphs,''
\newblock in {\em ICASSP}, 2021, pp. 5230--5234.

\bibitem{natali2021online}
A.~Natali, M.~Coutino, E.~Isufi, and G.~Leus,
\newblock ``Online time-varying topology identification via prediction-correction algorithms,''
\newblock in {\em ICASSP}, 2021, pp. 5400--5404.

\bibitem{zaman2020online}
B.~Zaman, L.~M.~L. Ramos, D.~Romero, and B.~Beferull-Lozano,
\newblock ``Online topology identification from vector autoregressive time series,''
\newblock {\em IEEE Transactions on Signal Processing}, vol. 69, pp. 210--225, 2020.

\bibitem{liu2019graph}
Y.~Liu, L.~Yang, K.~You, W.~Guo, and W.~Wang,
\newblock ``Graph learning based on spatiotemporal smoothness for time-varying graph signal,''
\newblock {\em IEEE Access}, vol. 7, pp. 62372--62386, 2019.

\bibitem{liu2019spatiotemporal}
Y.~Liu, L.~Yang, G.~Wenbin, T.~Peng, and W.~Wang,
\newblock ``Spatiotemporal smoothness-based graph learning method for sensor networks,''
\newblock in {\em WCNC}. IEEE, 2019, pp. 1--6.

\bibitem{yamada2019time}
K.~Yamada, Y.~Tanaka, and A.~Ortega,
\newblock ``Time-varying graph learning based on sparseness of temporal variation,''
\newblock in {\em ICASSP}, 2019, pp. 5411--5415.

\bibitem{Barbarossa_2020}
S.~Barbarossa and S.~Sardellitti,
\newblock ``Topological signal processing over simplicial complexes,''
\newblock {\em IEEE Transactions on Signal Processing}, vol. 68, pp. 2992--3007, 2020.

\bibitem{Barbarossa_2020_SPM}
S.~Barbarossa and S.~Sardellitti,
\newblock ``Topological signal processing: Making sense of data building on multiway relations,''
\newblock {\em IEEE Signal Processing Magazine}, vol. 37, no. 6, pp. 174--183, 2020.

\bibitem{DiLorenzo2018_sampling}
P.~Di~Lorenzo, P.~Banelli, E.~Isufi, S.~Barbarossa, and G.~Leus,
\newblock ``Adaptive graph signal processing: Algorithms and optimal sampling strategies,''
\newblock {\em IEEE Transactions on Signal Processing}, vol. 66, no. 13, pp. 3584--3598, 2018.

\bibitem{bib_GSP_filter_design}
N.~Tremblay, P.~Gonçalves, and P.~Borgnat,
\newblock ``Chapter 11 - design of graph filters and filterbanks,''
\newblock in {\em Cooperative and Graph Signal Processing}, P.~M. Djurić and C.~Richard, Eds., pp. 299--324. Academic Press, 2018.

\bibitem{Shuman_2011_Chebyshev}
D.~I. Shuman, P.~Vandergheynst, and P.~Frossard,
\newblock ``Chebyshev polynomial approximation for distributed signal processing,''
\newblock in {\em DCOSS}, 2011, pp. 1--8.

\bibitem{stankovic_2019_vertex}
L.~Stankovi{\'c} and E.~Sejdi{\'c},
\newblock {\em Vertex-frequency {A}nalysis of {G}raph {S}ignals},
\newblock Springer, 2019.

\bibitem{Chen_oversmoothing_2020}
D.~Chen, Y.~Lin, W.~Li, P.~Li, J.~Zhou, and X.~Sun,
\newblock ``Measuring and relieving the over-smoothing problem for graph neural networks from the topological view,''
\newblock {\em Proceedings of the AAAI}, vol. 34, pp. 3438--3445, Apr. 2020.

\bibitem{zhang2015graph}
C.~Zhang, D.~Flor{\^e}ncio, and P.~A. Chou,
\newblock ``Graph signal processing-a probabilistic framework,''
\newblock {\em Microsoft Res., Redmond, WA, USA, Tech. Rep. MSR-TR-2015-31}, 2015.

\bibitem{Kadambari_2022_Distributed}
S.~K. Kadambari, R.~Francis, and S.~P. Chepuri,
\newblock ``Distributed denoising over simplicial complexes using chebyshev polynomial approximation,''
\newblock in {\em EUSIPCO}, 2022, pp. 822--826.

\bibitem{nguyen2020_LMP}
N.~Nguyen, K.~Do{\u{g}}an{\c{c}}ay, and W.~Wang,
\newblock ``Adaptive estimation and sparse sampling for graph signals in alpha-stable noise,''
\newblock {\em Digital Signal Processing}, vol. 105, pp. 102782, 2020.

\bibitem{Spelta_2020_NLMS}
M.~J.~M. Spelta and W.~A. Martins,
\newblock ``Normalized lms algorithm and data-selective strategies for adaptive graph signal estimation,''
\newblock {\em Signal Processing}, vol. 167, pp. 107326, 2020.

\bibitem{Diniz_2007_adaptive_filtering}
P.~Diniz,
\newblock {\em Adaptive Filtering: Algorithms and Practical Implementation},
\newblock Springer, 01 2008.

\bibitem{yan_2022_sign}
Y.~Yan, E.~E. Kuruoglu, and M.~A. Altinkaya,
\newblock ``Adaptive sign algorithm for graph signal processing,''
\newblock {\em Signal Processing}, vol. 200, pp. 108662, 2022.

\bibitem{zhao_2024_sequential}
F.~Zhao and E.~E. Kuruoglu,
\newblock ``Sequential {Monte Carlo} graph convolutional network for dynamic brain connectivity,''
\newblock in {\em ICASSP}, 2024.

\bibitem{craddock2013neuro}
C.~Craddock, Y.~Benhajali, C.~Chu, F.~Chouinard, A.~Evans, A.~Jakab, B.~S. Khundrakpam, J.~D. Lewis, Q.~Li, M.~Milham, et~al.,
\newblock ``The neuro bureau preprocessing initiative: open sharing of preprocessed neuroimaging data and derivatives,''
\newblock {\em Frontiers in Neuroinformatics}, vol. 7, no. 27, pp. 5, 2013.

\bibitem{li2021braingnn}
X.~Li, Y.~Zhou, N.~Dvornek, M.~Zhang, S.~Gao, J.~Zhuang, D.~Scheinost, L.~H. Staib, P.~Ventola, and J.~S. Duncan,
\newblock ``Braingnn: Interpretable brain graph neural network for {fMRI} analysis,''
\newblock {\em Medical Image Analysis}, vol. 74, pp. 102233, 2021.

\bibitem{tumminello2005tool}
M.~Tumminello, T.~Aste, T.~Di~Matteo, and R.~N. Mantegna,
\newblock ``A tool for filtering information in complex systems,''
\newblock {\em Proceedings of the National Academy of Sciences}, vol. 102, no. 30, pp. 10421--10426, 2005.

\bibitem{zhu2021graph_heterophily}
J.~Zhu, R.~A. Rossi, A.~Rao, T.~Mai, N.~Lipka, N.~K. Ahmed, and D.~Koutra,
\newblock ``Graph neural networks with heterophily,''
\newblock in {\em Proceedings of the AAAI}, 2021, vol.~35, pp. 11168--11176.

\bibitem{2024_Stable_Muvunza}
T.~Muvunza, Y.~Li, and E.~E. Kuruoglu,
\newblock ``Stable probabilistic graphical models for systemic risk estimation,''
\newblock in {\em IEEE CAI}, 2024.

\bibitem{SAFFARIMIANDOAB2021115197}
F.~Saffarimiandoab, R.~Mattesini, W.~Fu, E.~E. Kuruoglu, and X.~Zhang,
\newblock ``Insights on features' contribution to desalination dynamics and capacity of capacitive deionization through machine learning study,''
\newblock {\em Desalination}, vol. 515, pp. 115197, 2021.

\bibitem{Isufi2019gvarma}
E.~Isufi, A.~Loukas, N.~Perraudin, and G.~Leus,
\newblock ``{Forecasting Time Series With VARMA Recursions on Graphs},''
\newblock {\em IEEE Transactions on Signal Processing}, vol. 67, no. 18, pp. 4870--4885, 2019.

\bibitem{francq2019garch}
C.~Francq and J.-M. Zakoian,
\newblock {\em GARCH models: structure, statistical inference and financial applications},
\newblock John Wiley \& Sons, 2019.

\bibitem{kalman_1960_new}
R.~E. Kalman,
\newblock ``A new approach to linear filtering and prediction problems,''
\newblock 1960.

\bibitem{Sagi_2023_graph_EFK_Kalman}
G.~Sagi, N.~Shlezinger, and T.~Routtenberg,
\newblock ``Extended {Kalman} filter for graph signals in nonlinear dynamic systems,''
\newblock in {\em ICASSP}, 2023, pp. 1--5.

\bibitem{alippi2023graph_kalman_new}
C.~Alippi and D.~Zambon,
\newblock ``Graph kalman filters,''
\newblock {\em arXiv preprint arXiv:2303.12021}, 2023.

\bibitem{costagli2007image}
M.~Costagli and E.~E. Kuruo{\u{g}}lu,
\newblock ``Image separation using particle filters,''
\newblock {\em Digital Signal Processing}, vol. 17, no. 5, pp. 935--946, 2007.

\bibitem{kipf2016semi}
T.~N. Kipf and M.~Welling,
\newblock ``Semi-supervised classification with graph convolutional networks,''
\newblock {\em arXiv preprint arXiv:1609.02907}, 2016.

\bibitem{kuruoglu1997new}
E.~Kuruoglu, C.~Molina, S.~Godsill, and W.~Fitzgerald,
\newblock ``A new analytic representation for the symmetric alpha-stable probability density function,''
\newblock in {\em Proceedings of the 5th World Meeting of the International Society for Bayesian Analysis (ISBA). ASA: American Statistical Association}, 1997, pp. 229--233.

\bibitem{herranz2004alpha}
D.~Herranz, E.~Kuruo{\u{g}}lu, and L.~Toffolatti,
\newblock ``An alpha-stable approach to the study of the p (d) distribution of unresolved point sources in cmb sky maps,''
\newblock {\em Astronomy \& Astrophysics}, vol. 424, no. 3, pp. 1081--1096, 2004.

\end{thebibliography}

\end{document}